\documentclass[12pt,usenatbib]{article}
\usepackage{graphicx}

\begin{document}
\newcommand{\ii}{\'{\i}}
\renewcommand{\thefootnote}{\alph{footnote}}

\def\lsim{\ ^{<}\!\!\!\!_{\sim}\>}
\def\gsim{\ ^{>}\!\!\!\!_{\sim}\>}

\centerline{\bf \Large Magnitude and size distribution of long-period}

\centerline{\bf \Large comets in Earth-crossing or approaching orbits}

\vspace{2cm}

\centerline{Julio A. Fern\'andez and Andrea Sosa}
\centerline{Departamento de Astronom\ii a, Facultad de Ciencias,}
\centerline{Igu\'a 4225, 11400 Montevideo, Uruguay}
\centerline{(email: julio@fisica.edu.uy, email: asosa@fisica.edu.uy)}

\vspace{2cm}


\vspace{4cm}


\vspace{6cm}

\centerline{{\large{\bf MNRAS}, in press}}

\vfill
\eject

\centerline{{\bf \large Abstract}}

\vspace{2cm}

We analyse the population of near-Earth Long-Period Comets (LPCs)
(perihelion distances $q < 1.3$ AU and orbital periods $P > 10^3$ yr).
We  have considered the sample of LPCs discovered during the period
1900-2009 and their estimated absolute total visual magnitudes $H$. For the
period 1900-1970 we have relied upon historical estimates of
absolute total magnitudes,
while for the more recent period 1970-2009 we have made our own
estimates of $H$ based on Green's photometric data base and IAU Circulars.
We have also used historical records for the sample of brightest
comets ($H < 4.5$) covering the period: 1500-1899, based mainly on
Vsekhsvyatskii, Hasegawa and Kronk catalogues. We find that the cumulative
distribution of $H$ can be represented by a three-modal law
of the form $\log_{10}N_{<H} = C + \alpha H$, where the $C'$s are
constants for the different legs, and $\alpha \simeq 0.28 \pm 0.10$
for $H < 4.0$, $\alpha \simeq 0.56 \pm 0.10$ for
$4.0 \leq H < 5.8$, and $\alpha \simeq 0.20 \pm 0.02$ for $5.8 \leq H <
8.6$. The large increase of the slope of the second leg of the
$H$-distribution might be at least partially attributed to splitting
of comet nuclei leading to the creation of two or more daughter
comets. The cumulative $H$-distribution tends to flatten for comets
fainter than $H \simeq 8.6$. LPCs
fainter than $H \simeq 12$ (or diametres $D \lsim 0.5$ km) are
extremely rare, despite several sky
surveys of near-Earth objects implemented during the last couple of
decades, suggesting a minimum
size for a LPC to remain active. We also find that about 30\% of
all LPCs with $q < 1.3$ AU are new (original bound energies $0 < E_{or}
< 10^{-4}$ AU$^{-1}$ ), and that among the new comets about half
come from the outer Oort cloud (energies $0 \lsim E_{or} \lsim 0.3
\times 10^{-4}$ AU$^{-1}$), and the other half from the inner Oort
cloud (energies $0.3 \times 10^{-4} \lsim E_{or} \lsim 10^{-4}$
AU$^{-1}$).

\bigskip


\vfill
\eject

\section{Introduction}

LPCs are natural probes to explore the comet reservoir in the outer
reaches of the solar system. Due to their great gaseous activity, even
small comets can be detected if they come close enough to the
Sun. Kres\'ak and Pittich (1978) estimated that about 60\% of all LPCs
in Earth-crossing orbits were being discovered by that time. As we
will see below, the discovery rate has increased to near completion,
at least for LPCs brighter than absolute magnitude $\sim 8.5$. The
degree of completeness falls sharply beyond Earth's orbit, as comets
become fainter because they are less active and are farther away from
Earth. Even when distant comets are discovered, it is very difficult
to predict their absolute brightness (i.e. measured ideally at 1 AU from
the Earth and from the Sun) because it depends on unreliable
extrapolations in heliocentric distance. Therefore, we have to look with
suspicion previous efforts  to try to derive the magnitude distribution
of LPCs based on samples containing distant comets (e.g. Hughes 1988, 2001).
We have thus decided to restrict our sample to comets with perihelion
distances $q < 1.3$ AU because it is more complete and because their
absolute magnitudes are obtained straightforward around $r \sim 1$ AU,
without needing to resort to uncertain large extrapolations.\\

Even though we have at present a rather good sky coverage that allows
us to detect most of the comets coming close to the Sun,
the computation of their masses or sizes remains as an extremely
difficult task. Sosa and Fern\'andez (2011)
have derived the masses of a sample of
LPCs from the estimated nongravitational forces that affect their
orbital motion. They also found a correlation of the masses or
sizes with the absolute total magnitude $H$, which can be expressed as

\begin{equation}
\log_{10}{R(\mbox{km})} = 0.9 - 0.13H,
\label{rvh}
\end{equation}
where $R$ is the radius of the comet nucleus, and we assume a mean
bulk density of 0.4 g cm$^{-3}$ for conversion
of masses to sizes. Equation (\ref{rvh}) will be very useful for our
goals since it will allow us to get a rough idea of the sizes and size
distribution of LPCs from the knowledge of their absolute total
magnitudes. A potential shortcoming of equation (\ref{rvh}) is that it
has been derived from a rather limited range of magnitudes: $H \sim
5-9$. We have then checked the validity of this equation for the
brightest LPC we have in our sample: C/1995 O1 (Hale-Bopp), for which
we find $H=-1.7$ (cf. Table \ref{sample_recent} below). Introducing
this value in equation (\ref{rvh}) we obtain $R=13.2$ km. Szab\'o et
al. (2011) have recently detected Hale-Bopp at 30.7 AU to the
Sun. From its observed magnitude and assuming a 4\% albedo, the
authors derive a radius of 60-65 km if the nucleus were inactive. Yet,
the authors suggest that some low-level activity may still be present,
so their estimated radius should be taken as an upper limit. In
conclusion, the computed $R$ value from equation (\ref{rvh}) for
Hale-Bopp may still be compatible with respect to its actual
value. This gives us some confidence for the use of this equation for
a range of $H$ wider than that from which it has been derived.\\

From the computation of masses and sizes, Sosa and Fern\'andez (2011)
have found that LPCs are hyper-active, i.e. with gas production rates
in general higher than those derived from thermal models of totally
free-sublimating surfaces of water ice. This might be explained as
the result of
frequent mini-outbursts and liberation of chunks of icy material that
quickly sublimate upon release, thus leading to erosion rates well
above those theoretically expected from a surface of water ice on a
free-sublimation regime. This agrees with copious evidence from the
connection between meteoroid streams and some short-period comets,
suggesting that the streams originate from discrete breakup events
and release of dust and small fragments, rather than from the normal
water ice sublimation (Jenniskens 2008). There are also many well
documented cases of LPCs on Earth-crossing orbits that disintegrated
during their passages as, for instance, comets C/1999 S4 (LINEAR),
C/2004 S1 (van Ness) (Sekanina et al. 2005), and C/2010 X1 (Elenin)
(see, e.g., Mattiazzo's (2011) report). At least two of them (C/1999
S4 and C/2010 X1) seem to be new, namely coming into the inner
planetary region for the first time (see Nakano
Notes\footnote{http://www.oaa.gr.jp/$\sim$oaacs/nk.htm} and Kinoshita's
electronic catalogue of comet
orbits\footnote{http://jcometobs.web.fc2.com/}
suggesting that small, faint
comets are not able to withstand a single perihelion
passage close to the Sun. There are also other comets observed
to split (e.g. Chen and Jewitt 1994, Sekanina 1997), thus creating
daughter comets that may last for several revolutions.\\

All the observed high activity and disintegration phenomena tells us
that comets could not last long in bound small-$q$ orbits so, either
they are dynamically ejected, or they fade away after a few passages,
at least those of typical kilometre-size. We will come back to this
problem when we try to estimate the fraction of new comets among the
LPC population.\\

The motivation of this paper is to rediscuss the magnitude distribution
of LPCs. We want to compare our derived cumulative $H$-distribution
with those from other authors (e.g. Everhart 1967b, Sekanina and
Yeomans 1984, Hughes 1988, 2001) and, in particular, to check if there
ia a knee at $H \sim 6$ at which the $H$-distribution passes from a steep
slope to a shallow one. Once
the magnitude distribution is derived, we will be able to determine
the size distribution by means of equation (\ref{rvh}), and to compare
it with the size distributions of other populations of primitive bodies.\\

The paper has been organised as follows. The second section describes
the chosen comet samples and the method developed to compute absolute
total magnitudes. The third section analyses the completeness of our
sample of discovered comets and potential observational biases. The
fourth section deals with the cumulative distribution of absolute
total magnitudes. The fifth section tries to answer the question: what
is the fraction of new comets within the sample of observed near-Earth
LPCs?. The sixth section
discusses the physical processes leading to erosion and fragmentation
of comet nuclei. The seventh section presents a simple numerical model
that combines physical and dynamical effects to try to explain the
main observed features of the cumulative magnitude distribution of
LPCs and the observed ratio new-to-evolved LPCs. Finally, the eighth
section summarises our main conclusions and results.

\section{The computed absolute total visual magnitudes of LPCs}
\label{method}

\subsection{The samples}

The samples adopted for photometric studies all
involve LPCs in Earth-approaching or crossing orbits (perihelion
distances $q < 1.3$ AU) for which we have a greater degree of
completeness and better photometric data. We have used the following
source of data:

\begin{itemize}
\item Ancient LPCs (1500-1899) brighter than $H=4.5$: we used as
references the catalogues of Vsekhsvyatskii (1964a), Hasegawa
(1980), and Kronk (1999, 2003). The sample of comets discovered
in the period 1650-1899 brighter than $H=4$, for which we have
more reliable photometric and orbit data, is shown in Table
\ref{sample_ancient}. It has been essentially extracted from
Vsekhsvyatskii's catalogue.

\item Modern LPCs (1900-1980): The magnitudes have been drawn from
Vsekhsvyatskii (1964a) and furhter updates (Vsekhsvyatskii 1963, 1964b,
1967, and Vsekhsvyatskii and Il'ichishina 1971), Whipple (1978),
Meisel and Morris (1976, 1982). The magnitudes are shown in Table
\ref{sample_modern}.

\item Recent LPCs (1970-2009) for which we made our own estimates of
  absolute magnitudes (see procedure below) based on Daniel Green's
data base of reported visual magnitudes and \emph{International
Astronomical Union Circulars} (IAUCs) reports. The magnitudes
are shown in Table \ref{sample_recent}.
\end{itemize}

\begin{table}[h]
\centering
\caption{{\small The estimated absolute total visual magnitudes of the
    brightest ($H$ $<$ 4.0) ancient LPCs (1650 - 1899).}}
\vspace{1mm}
\begin{scriptsize}
\begin{tabular}{|l r r r|l r r r|}
\hline
 Comet & $q$ & $i$ & $H$ & Comet & $q$  & $i$ & $H$  \\
 & (AU) & (deg) & & & (AU)& (deg) & \\ \hline \hline
1664  W1  & 1.026& 158.7&    2.4 & 1783  X1  & 0.708& 128.9&    3.6 \\
1672  E1  & 0.695& 83.0&    3.4 & 1807  R1 & 0.646& 63.2&     1.6   \\
1739  K1  & 0.674& 124.3&    3.3 & 1811  F1 & 1.035& 106.9&     0.0\\
1742  C1  & 0.766& 112.9&    3.9 & 1821  B1 & 0.092& 106.5&     3.4\\
1743  X1  & 0.222& 47.1&    0.5 &  1822  N1 & 1.145& 127.3&     3.0  \\
1760  B1  & 0.801& 79.1&    3.3 &  1825  N1 & 1.241& 146.4&     2.2\\
1762  K1  & 1.009& 85.7&    3.0 &  1858  L1 & 0.996& 116.9&     3.3  \\
1769  P1  & 0.123& 40.7&    3.2 & 1865  B1 & 0.026& 92.5&     3.8  \\
1773  T1  & 1.127& 61.2&    2.5 &  1892  E1 & 1.027& 38.7&     3.2 \\ \hline
\end{tabular}
\end{scriptsize}
\label{sample_ancient}
\end{table}

\begin{table}[h]
\centering
\caption{{\small The selected LPCs with historical estimates of $H$. The most uncertain values of $E_{or}$ are shown between brackets (corresponding to a value of 2B in the Marsden \& Williams's orbit determination quality code or a value of 5 in the Kinoshita's orbit determination quality code).}}
\vspace{1mm}
\begin{scriptsize}
\begin{tabular}{|l r r r r|l r r r r|}
\hline
 Comet & $q$ & $i$  & $E_{or}$  & $H$ & Comet & $q$  & $i$ & $E_{or}$  & $H$\\
 & (AU) & (deg) & ($\times 10^{-6}$ AU$^{-1}$)& & & (AU)& (deg) & ($\times 10^{-6}$ AU$^{-1}$) &\\ \hline \hline
1900 O1	& 1.015 & 62.5  & [610] &   8.6 & 1940 S1	& 1.062 & 133.1 & [-124]&  10.9\\
1901 G1	& 0.245 & 131.1 & -    &   5.9 & 1941 B2	& 0.790 & 168.2 &  2029&   6.0\\
1902 G1	& 0.444 & 65.2  & -    &  11.7 & 1941 K1	& 0.875 & 94.5 &  78&   6.9\\
1902 R1	& 0.401 & 156.3 &  27    &   6.2 & 1943 R1	& 0.758 & 161.3 & - &  11.0\\
1903 A1	& 0.411 & 30.9  &  1063    &   8.4 & 1943 W1	& 0.874 & 136.2 & - &  10.0\\
1903 H1	& 0.499 & 66.5  & -    &   9.0 & 1945 L1	& 0.998 & 156.5 & - &  10.4\\
1903 M1	& 0.330 & 85.0  &  33    &   6.4 & 1945 W1	& 0.194 & 49.5  & - &   9.6\\
1905 W1	& 1.052 & 140.6 & -    &   9.5 & 1946 K1	& 1.018 & 169.6 & - &   9.4\\
1905 X1	& 0.216 & 43.6  & -    &   8.3 & 1946 P1	& 1.136 & 57.0  &  44&   4.6\\
1906 B1	& 1.297 & 126.4 & -75    &   7.6 & 1947 F1	& 0.560 & 39.3  &  5924&   9.1\\
1906 F1	& 0.723 & 83.5  & -    &  10.2 & 1947 F2	& 0.962 & 129.1 & - &  11.2\\
1907 G1	& 0.924 & 110.1 & -    &  10.0 & 1947 S1	& 0.748 & 140.6 &  24&   6.5\\
1907 L2	& 0.512 & 8.9   &  2650  &   4.2 & 1947 V1	& 0.753 & 106.3 & - &   9.8\\
1907 T1	& 0.983 & 119.6 & -    &   9.1 & 1947 X1-A & 0.110& 138.5 & - &   6.2\\
1908 R1	& 0.945 & 140.2 &  174   &   4.1 & 1948 L1	& 0.208 & 23.1  & [525]&   8.0\\
1909 L1	& 0.843 & 52.1  & -    &  10.9 & 1948 V1	& 0.135 & 23.1  &  1294&   5.5\\
1910 A1	& 0.129 & 138.8 &  135   &   5.2 & 1948 W1	& 1.273 & 87.6  &  2633&   6.0\\
1911 N1 & 0.684 & 148.4 &  6337  &   7.6 & 1951 C1	& 0.719 & 87.9  & - &   9.7\\
1911 O1	& 0.489 & 33.8  &   6280 &   5.4 & 1951 P1	& 0.740 & 152.5 &  1348&   9.0\\
1911 S2	& 0.788 & 108.1 &  2491 &   6.4 & 1952 M1	& 1.202 & 45.6 &  148&   9.0\\
1911 S3	& 0.303 & 96.5  & -74    &   5.8 & 1952 W1	& 0.778 & 97.2 & -125&   8.8\\
1912 R1	& 0.716 & 79.8  &  45    &   6.2 & 1953 G1	& 1.022 & 93.9 &  2983&  11.1\\
1912 V1	& 1.107 & 124.6 & -    &   8.0 & 1953 T1	& 0.970 & 53.2 & - &   7.8\\
1913 Y1	& 1.104 & 68.0  &  29    &   1.5 & 1953 X1	& 0.072 & 13.6 & - &   5.9\\
1914 F1	& 1.199 & 23.9  &  126 &   9.6 & 1954 M2	& 0.746 & 88.5 &  36&   8.9\\
1914 J1	& 0.543 & 113.0 & -    &   8.2 & 1954 O1	& 0.677 & 116.2&  49&   7.3\\
1914 S1	& 0.713 & 77.8  & [2239] &  6.5 & 1955 O1	& 0.885 & 107.5& -727 &   6.8\\
1915 C1	& 1.005 & 54.8  & -    &   4.5 & 1956 E1	& 0.842 & 147.5& - &  10.5\\
1915 R1	& 0.443 & 53.5  & -    &  10.0 & 1956 R1	& 0.316 & 119.9& - &   5.4\\
1917 H1	& 0.764 & 158.7 & -    &  10.1 & 1957 P1	& 0.355 & 93.9 &  2001&   4.0\\
1918 L1	& 1.102 & 69.7  & -    &  10.0 & 1957 U1	& 0.539 & 156.7& - &  10.6\\
1919 Q2	& 1.115 & 46.4  &  20    &   4.6 & 1959 O1	& 1.250 & 12.8 & [-446]&  11.0\\
1919 Y1	& 0.298 & 123.2 & -    &  12.4 & 1959 Q1	& 1.150 & 48.3 &  593&   9.6\\
1920 X1	& 1.148 & 22.0  & [5023] & 11.9 & 1959 Q2	& 0.166 & 107.8& - &   9.5\\
1921 E1	& 1.008 & 132.2 &  18 &   6.8 & 1959 X1	& 1.253 & 19.6 &  69&   6.3\\
1922 W1	& 0.924 & 23.4  & -    &   7.5 & 1959 Y1	& 0.504 & 159.6& - &   8.6\\
1923 T1	& 0.778 & 113.8 & -    &  10.0 & 1960 B1	& 1.171 & 69.5 & - &  10.9\\
1924 R1	& 0.406 & 120.1 & -    &   7.5 & 1961 O1	& 0.040 & 24.2 & [792]&   8.0\\
1925 G1	& 1.109 & 100.0 &  40    &   5.5 & 1962 C1	& 0.031 & 65.0 &  25&   6.2\\
1925 V1	& 0.764 & 144.6 & -    &   9.7 & 1962 H1	& 0.653 & 72.9 & - &  10.4\\
1925 X1	& 0.323 & 123.0 & -    &   9.3 & 1964 L1	& 0.500 & 161.8&  8131&   8.5\\
1927 A1	& 1.036 & 92.4  & -    &   8.3 & 1964 P1	& 1.259 & 68.0 &  2721&   6.8\\
1927 B1	& 0.752 & 83.7  & -    &  11.3 & 1965 S2	& 1.294 & 65.0 & - &   9.3\\
1927 X1	& 0.176 & 85.1  &  1674  &   5.2 & 1966 R1	& 0.882 & 48.3 & - &   7.8\\
1929 Y1	& 0.672 & 124.5 & -    &   8.4 & 1966 T1	& 0.419 & 9.1&  49&  10.2\\
1930 D1	& 1.087 & 99.9  & -    &  12.5 & 1967 C1	& 0.457 & 106.5& - &  10.5\\
1930 L1	& 1.153 & 97.1  & -    &   8.8 & 1967 M1	& 0.178 & 56.7& - &   7.3\\
1931 P1	& 0.075 & 169.3 &  300   &   7.0 &	1968 H1	& 0.680 & 102.2& - &  11.0\\
1933 D1	& 1.001 & 86.7  & -    &   9.8 &	1968 L1	& 1.234 & 61.8 & - &  10.3\\
1936 K1	& 1.100 & 78.5  &  8294  &   6.8 &	1968 N1	& 1.160 & 143.2& -82&   5.5\\
1937 N1	& 0.863 & 46.4  &  124   &   6.1 &	1968 Q2	& 1.099 & 127.9& - &   6.9\\
1939 B1	& 0.716 & 63.5  & [7813] &   9.2 &	1969 P1 & 0.774 & 8.9  & - &   8.0\\
1939 H1	& 0.528 & 133.1 & -    &   7.1 &	1969 T1	& 0.473 & 75.8 &  507&   5.9\\
1939 V1	& 0.945 & 92.9  & [3327] &  10.2 &	1975 E1	& 1.217 & 55.2 &  23&   6.7\\
1940 R2	& 0.368 & 50.0  &  1   &   6.1 &	1978 T3 & 0.432 & 138.3& - &  11.1\\ \hline
\end{tabular}
\end{scriptsize}
\label{sample_modern}
\end{table}

\begin{table}[h]
\centering
\caption{{\small The selected LPCs with our estimates of $H$. The most uncertain values of $E_{or}$ are shown between brackets (corresponding to a value of 2B in the Marsden \& Williams's orbit determination quality code or a value of 5 in the Kinoshita's orbit determination quality code).}}
\vspace{1mm}
\begin{scriptsize}
\begin{tabular}{|l r r r r r |l r r r r r|}
\hline
 Comet & $q$ & $i$  & $E_{or}$  & $H$ & $QC$ & Comet & $q$  & $i$ & $E_{or}$  & $H$ & QC  \\
 &  (AU) & (deg) & ($\times 10^{-6}$ AU$^{-1}$)& & & & (AU)& (deg) & ($\times 10^{-6}$ AU$^{-1}$) & & \\ \hline \hline
	1969 Y1   	& 0.538 & 90.0 & - &  4.1    & B & 1995 Q1	  	& 0.436 & 147.4&  4458&  7.1 & B\\
	1970 B1   	& 0.066 & 100.2& - &  8.7	& D & 1995 Y1	  	& 1.055 & 54.5 & -58&  7.0 & B\\
	1970 N1	  	& 1.113 & 126.7&  283&  5.2  	& A & 1996 B2	   	& 0.230 & 124.9& 1508&  4.5 & A \\
	1970 U1	   	& 0.405 & 60.8 & - &  8.1  	& C & 1996 J1-B     & 1.298 & 22.5& -1&  7.4 & D \\
	1971 E1	   	& 1.233 & 109.7& 310&  5.5  	& C & 1996 N1	   	& 0.926 & 52.1& -161&  8.4 & A \\
	1972 E1	   	& 0.927 & 123.7& [2297]&  8.2  	& C & 1996 Q1	  	& 0.840 & 73.4 & [1826]&  6.9 & C\\
	1973 E1	   	& 0.142 & 14.3 &  20&  5.7  	& A & 1997 N1	   	& 0.396 & 86.0 & - &  9.7 & C\\ 	
	1974 C1	   	& 0.503 & 61.3 & 628&  7.6 	& B	& 1998 J1	   	& 0.153 & 62.9 & - &  5.8 & C\\
	1974 V2   	& 0.865 & 134.8& - &  8.9	& D & 1998 P1	  	& 1.146 & 145.7& 222&  6.5 & A\\
	1975 N1	  	& 0.426 & 80.8 &  817&  6.7  	& A & 1999 A1	   	& 0.731 & 89.5 & [6186]& 11.5 & C\\
	1975 V1-A  	& 0.197 & 43.1 &  1569&  5.6  	& B & 1999 H1	   	& 0.708 & 149.4&  1313&  6.1 & A\\
	1975 V2	   	& 0.219 & 70.6 & -56&  8.7	& C	& 1999 J3	   	& 0.977 & 101.7&  1150&  8.2 & A\\
	1975 X1	   	& 0.864 & 94.0 & [-734]&  11.6	& D	& 1999 N2	   	& 0.761 & 111.7&  3442&  7.9 & C\\ 	
	1976 E1  	& 0.678 & 147.8& - &  11.4   & D	& 1999 S4	   	& 0.765 & 149.4& 2&  7.7 & C\\
	1977 H1   	& 1.118 & 43.2 & - &  12.2   & D	& 1999 T1	   	& 1.172 & 80.0 & 1147&  5.7 & B\\
	1977 R1	   	& 0.991 & 48.3 &  231&  6.4	& A	& 2000 S5       & 0.602 & 53.8 & - & 10.2 & D\\
	1978 C1	   	& 0.437 & 51.1 & - &  6.8    & C	& 2000 W1	   	& 0.321 & 160.2& -7& 10.2 & B\\
	1978 H1	   	& 1.137 & 43.8 &  24&  3.6	& C	& 2000 WM1      & 0.555 & 72.6 & 522&  6.4 & B\\
	1978 T1	  	& 0.370 & 67.8 & [5245]&  7.7	& B	& 2001 A2-A	   	& 0.779 & 36.5 & 1112&  7.2 & A\\
	1979 M1	  	& 0.413 & 136.2&  33&  11.5	& C	& 2001 Q4	   	& 0.962 & 99.6 & - &  5.1 & A\\
	1980 O1	  	& 0.523 & 49.1 & - &  8.0	& C	& 2002 F1	  	& 0.438 & 80.9 & 1284&  8.5 & B\\
	1980 Y1	   	& 0.260 & 138.6&  964&  6.9	& C	& 2002 O4	   	& 0.776 & 73.1 & -772&  7.9 & C\\
	1982 M1	   	& 0.648 & 84.5 &  1666&  7.3	& A	& 2002 O6	   	& 0.495 & 58.6 & - &  9.5 & C\\
	1983 J1	   	& 0.471 & 96.6 & [378]&  11.0	& C	& 2002 O7	   	& 0.903 & 98.7 &  27&  9.7 & C\\
	1984 N1	  	& 0.291 & 164.2&  510&  7.5	& B	& 2002 Q5      	& 1.243 & 149.2&  58& 12.0 & D\\
	1984 S1	   	& 0.857 & 145.6& - &  11.6	& D	& 2002 T7	  	& 0.615 & 160.6& 13&  4.6 & B\\
	1984 V1	   	& 0.918 & 65.7 &  1299&  8.1	& A	& 2002 U2     	& 1.209 & 59.1 &  1075& 12.2 & D\\
	1985 K1	  	& 0.106 & 16.3 & - &  8.5	& C	& 2002 V1	   	& 0.099 & 81.7 &  2297&  6.2 & A\\
	1985 R1	  	& 0.695 & 79.9 &  558&  7.8	& A	& 2002 X5	   	& 0.190 & 94.2 & 879&  7.0 & B\\
	1986 P1-A	& 1.120 & 147.1& 32&  5.0	& B	& 2002 Y1	   	& 0.714 & 103.8&  4102&  6.5 & A\\
	1987 A1	   	& 0.921 & 96.6 & -121&  9.8	& C	& 2003 K4	  	& 1.024 & 134.3&  23&  4.7 & B\\
	1987 B1	   	& 0.870 & 172.2&  5034&  6.2	& A	& 2003 T4	   	& 0.850 & 86.8 & -1373&  7.4 & B\\
	1987 P1	   	& 0.869 & 34.1 &  6380&  5.4	& A	& 2004 F4	   	& 0.168 & 63.2 &  5164&  8.1 & C\\
	1987 Q1	   	& 0.603 & 114.9&  526&  8.3	& A	& 2004 G1    	& 1.202 & 114.5 & - & 13.0 & D\\
	1987 T1	  	& 0.515 & 62.5 & - &  8.0	& C	& 2004 H6	   	& 0.776 & 107.7 & -124&  6.9 & C\\
	1987 U3	   	& 0.841 & 97.1 &  1491&  5.5	& B	& 2004 Q2	   	& 1.205 & 38.6 & 407&  4.9 & B\\
	1987 W1	  	& 0.199 & 41.6 & - &  9.4	& C	& 2004 R2	   	& 0.113 & 63.2 & - &  9.6 & C\\
	1988 A1	  	& 0.841 & 73.3 &  4881 &  5.5	& A	& 2004 S1     	& 0.682 & 114.7& - &  12.5 & D\\
	1988 F1   	& 1.174 & 62.8 &  1725 &  7.3    & D	& 2004 V13      & 0.181 & 34.2 & - & 13.7 & D\\
	1988 J1   	& 1.174 & 62.8 &  1725 &  8.2 	& D & 2005 A1-A	   	& 0.907 & 74.9 &  94&  7.8 & B\\
	1988 P1	   	& 0.165 & 40.2 & - &  7.7	& C	& 2005 K2-A	  	& 0.545 & 102.0 & - & 13.4 & C\\
	1988 Y1	  	& 0.428 & 71.0 & - &  12.4	& C	& 2005 N1	  	& 1.125 & 51.2 &  1289&  9.7 & C\\
	1989 Q1	  	& 0.642 & 90.1 & 91&  7.2	& A	& 2006 A1	  	& 0.555 & 92.7 &  783 &  7.4 & B\\
	1989 T1	  	& 1.047 & 46.0 &  9529 &  10.0	& B	& 2006 M4	  	& 0.783 & 111.8 &  207&  5.6 & C\\
	1989 W1	   	& 0.301 & 88.4 &  658 &  7.6	& A	& 2006 P1	  	& 0.171 & 77.8 &  37&  3.9 & B\\
	1989 X1	   	& 0.350 & 59.0 &  32 &  6.2	& B	& 2006 VZ13     & 1.015 & 134.8 &  14&  8.0 & D\\
	1990 E1	   	& 1.068 & 48.1 & - &  6.3	& C	& 2006 WD4      & 0.591 & 152.7 &  2247& 14.0 & D\\
	1990 K1	   	& 0.939 & 131.6& -58&  4.6	& A	& 2007 E2	   	& 1.092 & 95.9 &  970&  8.5 & C\\
	1990 N1	   	& 1.092 & 143.8&  4692 &  5.0	& B	& 2007 F1	  	& 0.402 & 116.1 & 823&  8.2 & B\\
	1991 T2	   	& 0.836 & 113.5&  936 &  7.7	& B	& 2007 N3	  	& 1.212 & 178.4 &  30&  5.3 & B\\
	1991 X2	   	& 0.199 & 95.6 & [57]&  10.2	& C	& 2007 P1       & 0.514 & 118.9 & - & 12.0 & D\\
	1991 Y1	   	& 0.644 & 50.0 &  -94&  9.5	& B	& 2007 T1	   	& 0.969 & 117.6 &  627&  7.7 & B\\
	1992 B1	   	& 0.500 & 20.2 & - &  10.4	& D	& 2007 W1	  	& 0.850 & 9.9 &  -10&  8.2 & A\\
	1992 F1	   	& 1.261 & 79.3 &  3422 &  5.6	& C	& 2008 A1	  	& 1.073 & 82.5 &  96&  5.6 & B\\
	1992 J2   	& 0.592 & 158.6& - &  11.0	& D & 2008 C1	   	& 1.262 & 61.8 &  17&  7.0 & C\\
	1992 N1	   	& 0.819 & 57.6 & - &  8.4	& D	& 2008 J4      	& 0.447 & 87.5& - & 14.3 & D\\
	1993 Q1	   	& 0.967 & 105.0&  3&  6.7	& B	& 2008 T2	  	& 1.202 & 56.3 & 8&  6.3 & B\\
	1993 Y1	  	& 0.868 & 51.6 & - &  8.3	& A	& 2009 F6       & 1.274 & 85.8 & 1421&  6.1 & D\\
	1994 E2   	& 1.159 & 131.3&  2681&  12.1   & D & 2009 G1	& 1.129 & 108.3& - &  9.0 & D\\
	1994 N1	  	& 1.140 & 94.4 &  1324&  7.7	& B	& 2009 O2   & 0.695 & 108.0 & 325&  8.8 & D\\
	1995 O1	   	& 0.914 & 89.4 &  3800&  -1.7	& A	& 2009 R1	& 0.405 & 77.0 & 13&  6.6 & D\\ \hline
\end{tabular}
\end{scriptsize}
\label{sample_recent}
\end{table}

\subsection{The method} \label{method}

The precise determination of comet total magnitudes is an elusive
problem since active comets do not appear as point-like sources but as
nebulosities. Aperture effects are among the several causes that
can make observers to underestimate the comet magnitudes. CCD
estimates are found to be, in general, much fainter than the visual
estimates (i.e. those made visually by telescope, binoculars or naked
eye). From a small set of LPCs that have both visual and CCD
magnitudes, taken at about the same time, we found that the CCD
magnitudes are on average about 1.5 magnitudes fainter than the
corresponding total visual magnitudes.  Therefore, in our study we
have used in the overwhelming majority of cases only visual
estimates. Only for a few poorly observed comets we had to resort to
CCD magnitudes. The absolute total magnitude $H$ can be
determined by means of

\begin{equation}
m_h = H + 2.5 n \log_{10} r,
\label{mag_absoluta}
\end{equation}

\noindent where $m_h$ is the heliocentric total magnitude (i.e. the
apparent total magnitude $m$ corrected by the geocentric distance
$\Delta$, $m_h = m - 5\log_{10}\Delta$), $r$ is the heliocentric
distance in AU, and $n$ is known as the \emph{photometric index}.
If $n$ = 4 is assumed, the standard total magnitude
$H_{10}$ is determined instead of $H$. The visual apparent magnitudes
estimates were obtained from the \emph{International Comet Quaterly
  (ICQ)} archive (except those observations prior to 2006, which were
provided by Daniel W. Green), and from the IAUCs. We follow the procedure
explained in Sosa and Fern\'andez (2009, 2011) to extract and
reduce the observational data.\\

To determine $H$ and $n$ for a given comet from equation
(\ref{mag_absoluta}), we made a least-square linear fit between the
estimated $m_h$ values evaluated at the observational times $t$
and the logarithms of the computed distances $r(t)$. For some
comets the observational coverage was not good enough to properly
define $H$ from the linear fit (e.g. because of a lack of observations
around $r$ $\sim$ 1 AU, or because the photometric slope significantly
varies within the observed range of $r$). In addition, most comets
present a somewhat different slope before and after perihelion, hence
different pre-perihelion, post-perihelion, and combined linear
fits were made in such cases. Therefore, for each comet of the studied
sample, we made an educated guess to determine $H$ (based not only on
the linear fits, but also taking into account the quality and
completeness of the light curve, as well as the observations closer to
$r$ $\sim$ 1 AU, when they existed). The results are presented in Table
\ref{sample_recent}. For a certain number of comets it was not possible
to fit equation (\ref{mag_absoluta}), not even estimate $H$ from $m_h$
estimates around $r$ $\sim$ 1 AU, due to the poor observational
data. For these comets we estimated $H_{10}$ instead of $H$ from the
scarce observations lying relatively far from $r$ $\sim$ 1 AU. For
a few of these comets that lacked visual observations, we had to
estimate $m_h$ from CCD observations by using the
empirical relation (cf. above): $m_{vis} = m_{CCD} - 1.5$. After the
processing of the observational data,
we were able to estimate $H$ for 122 LPCs of the selected sample. The
results are presented in Table \ref{sample_recent}.\\

\begin{figure}[h]
\centerline{\includegraphics[width=10cm]{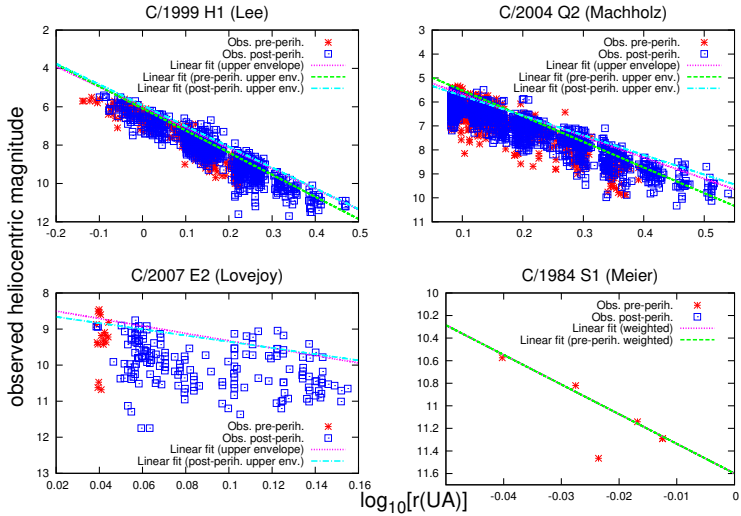}}
\caption{Heliocentric total visual magnitudes as a function of the
  logarithm of the heliocentric distance, for comets with light curves
  of different quality classes: C/1999 H1 (Lee) $QC$ = $A$ (upper left
  panel); C/2004 Q2 (Machholz) $QC$ = $B$ (upper right
  panel); C/2007 E2 (Lovejoy) $QC$ = $C$ (lower left panel); and
  C/1984 S1 (Meier) $QC$ = $D$ (lower right panel).}
\label{QC}
\end{figure}

In order to assess the uncertainty of our $H$ estimates, we define four
quality classes (hereafter $QC$) for the studied comets, according to
the features of their respective light curves: the {\it good} quality
class ($QC$ = $A$) is assigned to those LPCs which present a good
photometric coverage, i.e. a relative large number of visual
observations covering a range of heliocentric distances of at least about
several tenths of AU, which also includes observations close to 1 AU
from the Sun, and enough pre- and post-perihelion observations to
define good pre-perihelion and post-perihelion fits to the light
curve. Also, a good or a rather good convergence between
the different fits (the pre-perihelion, the post-perihelion and the
overall fits) around $r = 1$ AU is also required for a comet's light
curve to be qualified as an $A$ type. We estimate an uncertainty $\Delta
H$ $\lsim$ 0.5 for our higher quality class comets. We define a {\it fair}
quality class ($QC$ = $B$) for those comets with a good number of
visual observations but that do not fulfill one or more requirements of the
$A$ class, because of a poor convergence of the pre-perihelion and the
post-perihelion light curve fits, or because a pre-perihelion (or a
post-perihelion) linear fit was not possible, or because it is necessary
to extrapolate by some hundredths AU to estimate $H$, or because
the comet exhibits a somewhat non-smooth photometric behavior (e.g. a
small outburst), slightly departing from a linear fit in the $m_h$
vs. $\log (r)$ domain. We estimate an uncertainty 0.5 $\lsim$ $\Delta H$
$\lsim$ 1.0 for the $B$ class. We define a {\it poor} quality class
($QC$ = $C$) when the number or the heliocentric distance range of the
observations are not good enough to properly define a linear fit to any
branch, or when although a linear fit to at least one of the branches
can be determined, it is
necessary to extrapolate by several tenths of AU to estimate $H$, or
because of a lack of observations around $r$ = 1 AU for both branches,
or because the comet exhibits a non-smooth photometric behavior (e.g
an outburst). We estimate an uncertainty $1.0 \lsim \Delta H \lsim
1.5$ for the $C$ class. Finally, we define a {\it very poor} quality class
($QC$ = $D$) for those comets with an insufficient number of
observations for which an analytic extrapolation (assuming a
photometric index of $n$ = 4) was needed to estimate the absolute
total magnitude. Besides, in some of these cases it was necessary to
convert CCD magnitudes to visual magnitudes as explained
above. These are the most uncertain estimates of $H$, which may be
$\Delta H \gsim 1.5$. The $QC$ code assigned to each comet of the
studied sample is shown in Table \ref{sample_recent}. Examples of
light curves of $QCs$ $A$, $B$, $C$, $D$ are shown in Fig. \ref{QC}.
The plots for the remaining comet light curves of our sample covering
the period 1970-2009 can be seen in
http://www.astronomia.edu.uy/depto/material/comets/.

\subsection{Comparison between our magnitude estimates and previous
  ones}

\begin{figure}[h]
\centerline{\includegraphics[width=12cm]{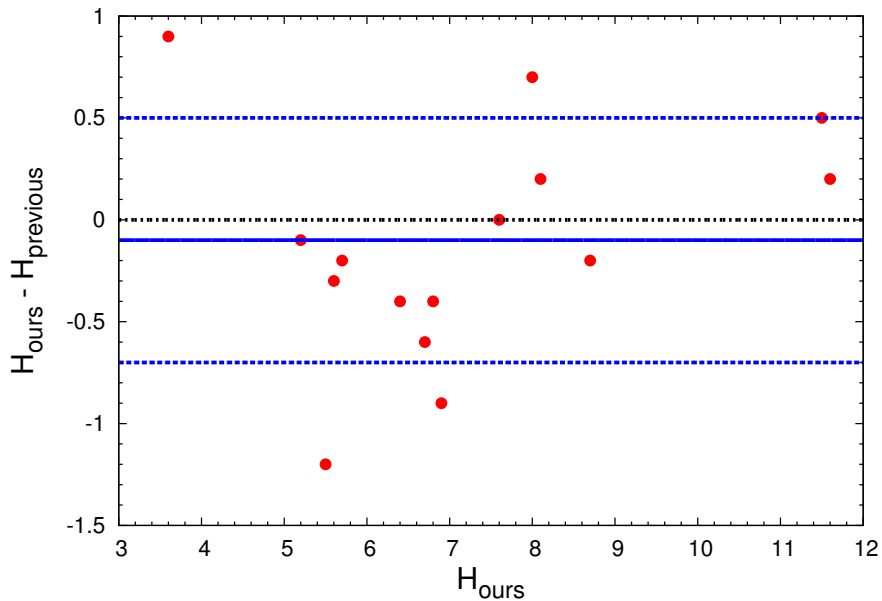}}
\caption{The difference between our estimated absolute total magnitude
  and previous estimates for a sample of fifteen comets observed
  between 1970 and 1980, as a function of our estimated magnitudes
$H_{ours}$. The horizontal lines represent the mean of the differences,
and the mean $\pm$ 1 $\sigma$. The plotted comets are: C/1970 N1,
C/1970 U1, C/1971 E1, C/1973 E1, C/1974 C1, C/1975 N1, C/1975 V1-A,
C/1975 V2, C/1975 X1, C/1977 R1, C/1978 C1, C/1978 H1, C/1979 M1,
C/1980 O1, C/1980 Y1.}
\label{difH}
\end{figure}

As a check to evaluate how our estimated absolute total magnitudes
compare with previous determinations, we used a set of comets
observed during the 1970s for which we have both, our own
estimates and previous ones, essentially from Vsekhsvyatskii, Whipple,
Meisel, and Morris (loc. cit.). As we can see in Fig. \ref{difH}, the
mean value of the differences is close
to zero, and only four comets (from a sample size of fifteen)
present differences larger than 1 $\sigma$ (i.e. differences between about
0.5 and 1.5 magnitudes). We then conclude that our estimates are consistent
with those from previous authors, which make us confident that we are not
introducing a significant bias in the $H$ estimates, when we combine those
from our comet sample for the period 1970-2009 with estimates from other
authors for older comet samples.


\section{The discovery rate}

\begin{figure}[h]
\centerline{\includegraphics[width=7cm]{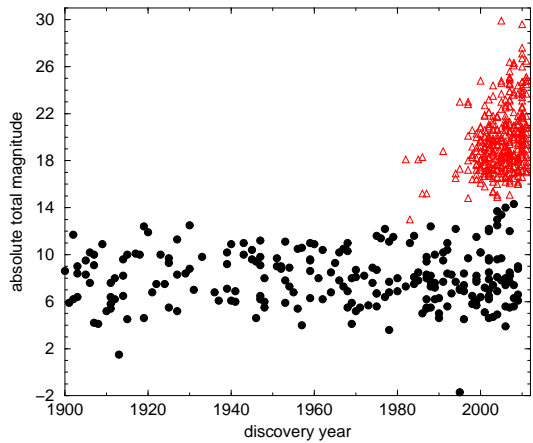}}
\caption{The discovery year versus absolute magnitudes of the sample
  of 232 LPCs (filled circles) and 375 NEAs in ``cometary'' orbits
  (aphelion distances $Q > 4.5$ AU) (red triangles). Since NEAs are
  inactive bodies, ``total'' means nuclear magnitude in their case.}
\label{modern_dy}
\end{figure}

We plot in Fig. \ref{modern_dy} the discovery year of LPCs with $q < 1.3$
AU for the period 1900-2009 versus their absolute total
magnitudes. We see that most magnitudes are below $H \simeq 12$, and
that this ceiling has changed very little through time despite the
systematic sky surveys implemented during the last two decades. By
contrast, these surveys have greatly contributed to a dramatic increase
in the discovery rate of Near-Earth Asteroids (NEAs) in ``cometary'' orbits
(i.e. with aphelion distances $Q > 4.5$ AU) fainter than absolute
magnitude 14, with a few as faint as magnitudes 25-30. This is in
agreement with the conclusion reached by Francis (2005), who found very
few LPCs with $H > 11$ from the analysis of a more restricted -but
more selected- sample of LPCs discovered by LINEAR that reached perihelion
between 2000 January 1 and 2002 December 31.\\

We also note in Fig. \ref{modern_dy} that the density of points for
the discovered comets tends to increase somewhat with time. We
actually note three regions: the less dense part for the period
1900-1944, an intermediate zone for 1945-1984, and the most dense part
for 1985-2009. The fact that the increase has been only very
moderate for the last century, and that it has been kept more or less
constant for the last 25 years, suggests us that the discovery rate
has attained near completion, at least for magnitudes $H \lsim 9$. We
have also investigated possible observation selection effects that may
have affected, or are still affecting, the discovery rate. We will
next analyse this point.

\subsection{The Holetschek effect}

\begin{figure}[h]
\centerline{\includegraphics[width=7cm]{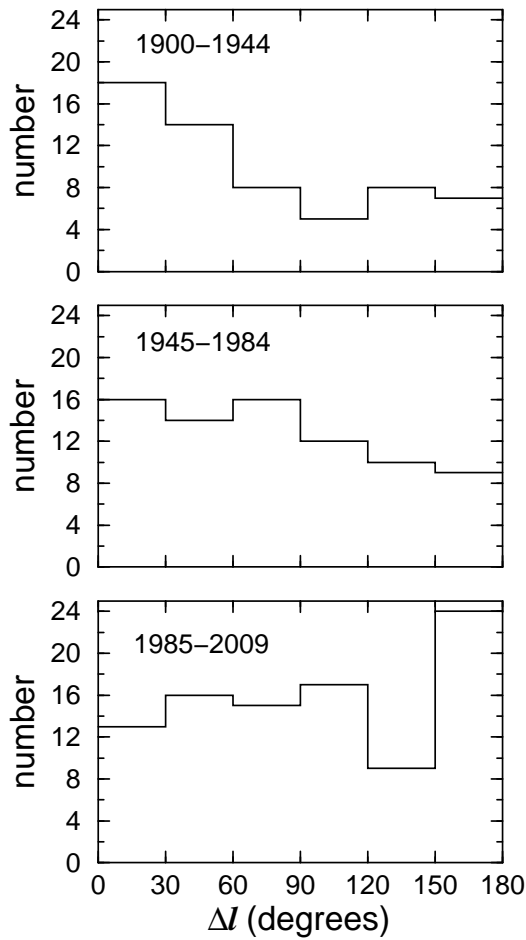}}
\caption{Distributions of longitude differences $\Delta l = l_c -
l_{\oplus}$, where $l_c$ is the heliocentric longitude of the comet
at perihelion,
and $l_{\oplus}$ that of the Earth at that time, for the LPC samples
for the periods indicated on the panels. To smooth out the histograms,
the half portions between 180$^{\circ}$ and 360$^{\circ}$ have been
folded over the first half portions between 0$^{\circ}$ and
180$^{\circ}$. A decrease in the number of comets reaching perihelion
near $\Delta l \simeq 180^{\circ}$ is noticed in the two older samples
(Holetschek effect).}
\label{holetschek}
\end{figure}

 The potential discovery of LPCs (i.e. comets that have been recorded only
 once during the age of scientific observation) is a function of its
 brightness (which depends on the perihelion distance) and the
 comet-Earth-Sun geometry. The Holetschek effect is the best known one
 (e.g. Everhart 1967a, Kres\'ak 1975), and is associated with the fact that
 comets reaching perihelion on the opposite side of the Sun, as seen from
the Earth, are less likely to be discovered. This effect essentially
affects comets in Earth-approaching or crossing orbits, as is the case
of our sample. In Fig. \ref{holetschek} we show the
differences in heliocentric longitude $\Delta l$ between the comet and
 the Earth computed at the time of the comet's perihelion passage (the
 ephemeris data were obtained from JPL Horizon's orbital
 integrator). We can see that the Holetschek effect is important for
 comets observed between 1900 and 1944; it is less important for comets
 observed between 1945 and 1984, and negligible for comets observed
 between 1985 and 2009, i.e. when dedicated surveys with CCD detectors
 began to operate. Hence, we consider the subsample of LPCs
 observed between 1985 and 2009 as an unbiased sample, at least as
 regards to this effect.

\subsection{The Northern-Southern asymmetry}

We have also investigated if it was a dominance of northern discoveries
against southern ones. The distribution of the sine of the comet's
declination $\delta$ at discovery does not show a significant drop for high
southern declinations ($\delta$ $<$ -30$^{\circ}$), hence we
conclude that the unequal coverage of the northern and southern
hemispheres has had little effect on comet discovery, at least for LPCs
with $q < 1.3$ AU discovered during the last century.

\section{The cumulative distribution of absolute visual magnitudes}

\subsection{The sample of LPCs for the period 1900-2009}

\begin{figure}[h]
\centerline{\includegraphics[width=14cm]{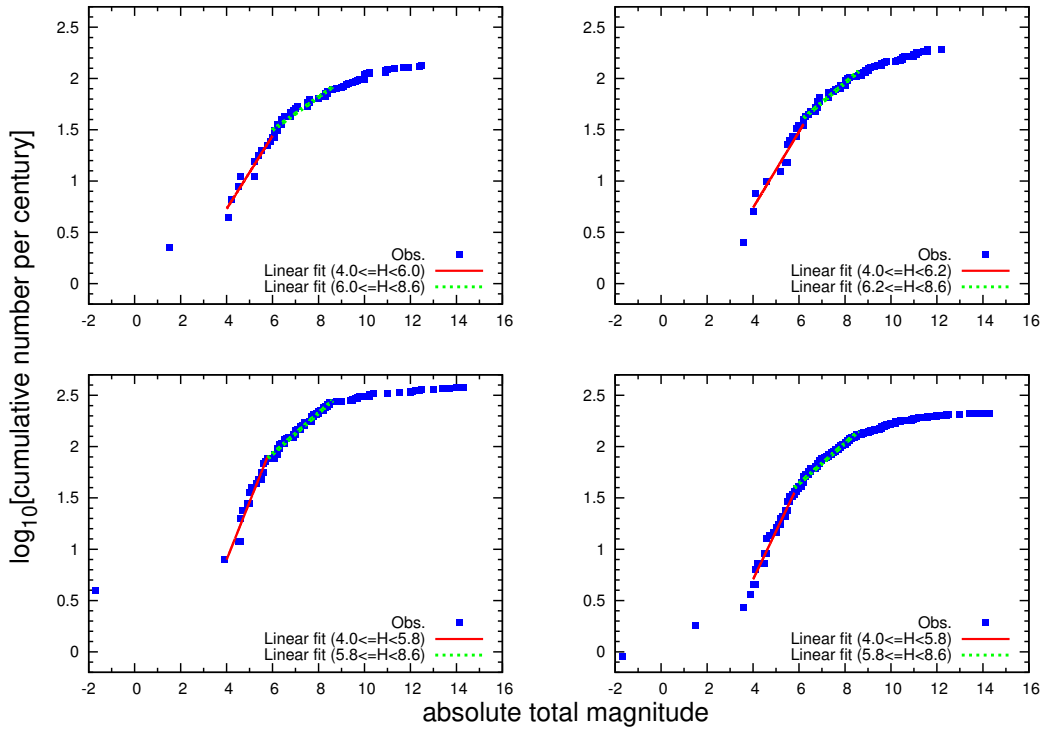}}
\caption{The cumulative distributions of H for the samples: 1900-1944
  (upper left panel), 1945-1984 (upper right panel), 1985-2009 (lower
  left panel), and the overall sample 1900-2009 (lower right panel).}
\label{mosaico}
\end{figure}

In Fig. \ref{mosaico} we present the logarithm of the cumulative number
of comets $N_{<H}$ having absolute total visual magnitudes smaller than
a specific value $H$, plotted as a function of the absolute magnitude, for
the overall comet sample 1900-2009, and for the three sub-samples: 1900-1944,
1945-1984, and 1985-2009. We have normalized the number of discovered
comets within the different periods to comets century$^{-1}$.
We found that within a certain range of $H$,
$\log_{10}N_{<H}$ could be well fitted by a linear relation, namely

\begin{equation}
\log_{10} N_{<H} \ = \ \alpha H + C,
\label{cumh}
\end{equation}

\noindent where $C$ is a constant, and the slope was found to be
$\alpha = 0.56 \pm 0.10$ for comets with $4.0 \leq H <5.8$, and
$\alpha = 0.20 \pm 0.02$ for comets with $5.8 \leq H < 8.6$, as inferred
from what we consider as the most unbiased sub-sample: LPCs for 1985-2009
(see lower left panel of Fig. \ref{mosaico}). We note that we used all
the observed comets for the period 1985-2009 for deriving the slopes of
equation (\ref{cumh}), including the most uncertain quality class D. We
then checked the previous results by considering only the quality
classes A, B and C, leaving aside D-quality comets. We found very minor
changes, of a couple of hundredths units in the slopes at most, so we
decided to keep the results for the complete sample.\\

We found similar
behaviours for the other sub-samples, namely a steep slope up to $H \sim
6$, and then a smooth slope up to $H \sim 8.6$. Yet, the derived values
are somewhat lower: $\alpha \sim 0.35-0.47$ for the first leg, and
$\alpha \sim 0.16-0.19$ for the second one, but we should bear in mind
that these sub-samples are presumably incomplete, thus affecting the
computed values of $\alpha$. We may further argue that because fainter
comets are more likely to be missed than brighter ones, the biased
cumulative magnitude distributions may be flatter than the real ones,
thus explaining the lower values of $\alpha$ computed for the older
sub-samples.

\subsection{The ancient comets}

\begin{figure}[h]
\centerline{\includegraphics[width=7cm]{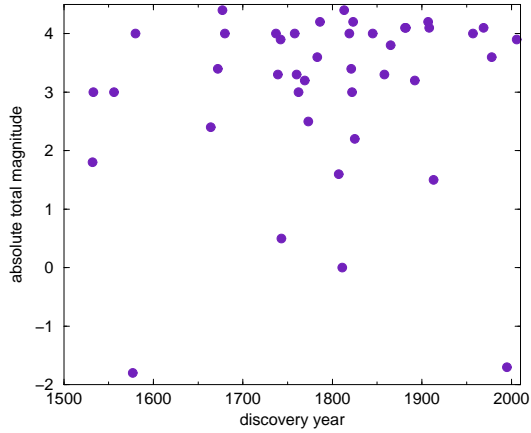}}
\caption{The discovery year versus absolute magnitudes of the sample
  of 34 ancient LPCs (1500-1899) plus 8 modern LPCs (1900-2009)
  brighter than $H = 4.5$.}
\label{ancient_dy}
\end{figure}

An inspection of the overall sample of Fig. \ref{mosaico} (lower right
panel) suggests us that the cumulative distribution of comets brighter
than $H \simeq 4$ tends to flatten, in other words, it seems to be
more bright comets than expected from the extrapolation to brighter
magnitudes of the steep slope found for magnitudes $4.0 \leq H < 5.8$.
Unfortunately the number of comets with $H < 4$ observed during
1900-2009 is too low to draw firm conclusions. To try to advance in
our knowledge of the brighter end of the magnitude distribution, we had
to resort to a comet sample observed over a longer time span. We then
assembled a sample of LPCs brighter than $H = 4.5$ observed during
1500-1900. Our main source
was Vsekhsvyatskii's (1964a) catalogue, complemented with information
provided by Kronk and Hasegawa. Even though we may consider the
photometric data of ancient comets of lower quality, as compared to
those for modern comets, for the time being it is the only source of
information available, and we hope from this to gain insight into the
question of what is the magnitude distribution of the brighter
comets. Fig. \ref{ancient_dy} shows the discovery rate of LPCs with $q
< 1.3$ AU brighter than $H = 4.5$ discovered over the period 1500-2009.
We observe a rather constant flux, at least from about 1650 up to the
present (that roughly corresponds to the telescopic era when photometric
observations became more rigorous), which suggests that the degree
of completeness of the discovery record of bright comets has been very
high since then.\\

\begin{figure}[h]
\centerline{\includegraphics[width=7cm]{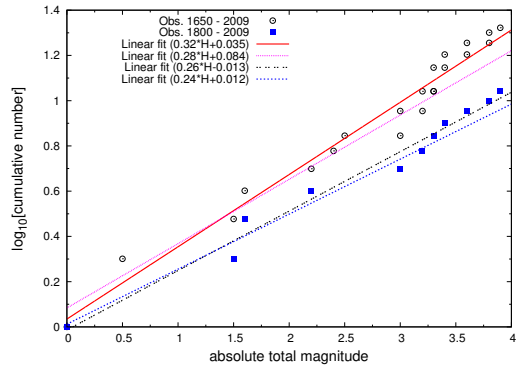}}
\caption{Cumulative distributions of $H$ for several subsets of
  the brightest comets $H < 4$ observed during the period
  1650-2009. The brightest comet of the sample, C/1995 O1 (Hale-Bopp),
  was not taken into account in the linear fits shown here.}
\label{ancient_cum}
\end{figure}

Fig. \ref{ancient_cum} shows the cumulative $H$-distribution of comets
brighter than $H=4$. To check how sensitive is the linear fit to
small changes in the sample, we tried to fit different sub-samples (as
shown in the figure). The samples of Fig. \ref{ancient_cum} did not
include the brightest comet (Hale-Bopp with $H=-1.7$), since its
detached position at the brightest end gives it a too strong weight in
the computed slope. For the two considered samples: 1650-2009 and
1800-2009, we analised two linear fits: one going up to a magnitude $H=4$,
and the other to only $H=3.2$. For $H=4$ we obtained computed slopes 0.32
and 0.26 for the samples 1650-2009 and 1800-2900, respectively, while
for $H=3.2$ the respective values decrease somewhat to 0.28 and 0.24,
respectively. The slight increase in the computed slope
as we pass from a limit at $H=3.2$ to $H=4$ may be explained as due to
the approach to the knee found at $H \sim 4$ and the transit to a much
steeper slope for fainter comets. We also find that the computed
slopes for the most restricted sample
1800-2009 are somewhat higher than the ones obtained for the whole
sample 1650-2009, which may de due to the greater incompleteness
of the older sample for 1650-1800.\\

We have also checked the
robustness of our computed results by considering two extreme cases:
one that includes the brightest comet Hale-Bopp, and the second one that
removes the two brightest comets (Hale-Bopp and C/1811 F1 with
$H=0$). In the first case we get values for the slope in the range
0.17-0.24; for the second we get values between 0.33-0.38.
As a conclusion, from the analysis of different sub-samples, that
contemplate different ranges of $H$ and two periods of time, we can
derive an average slope aroud 0.28 with an estimated uncertainty $\pm
0.1$.

\subsection{The overall sample}

\begin{figure}[h]
\centerline{\includegraphics[width=14cm]{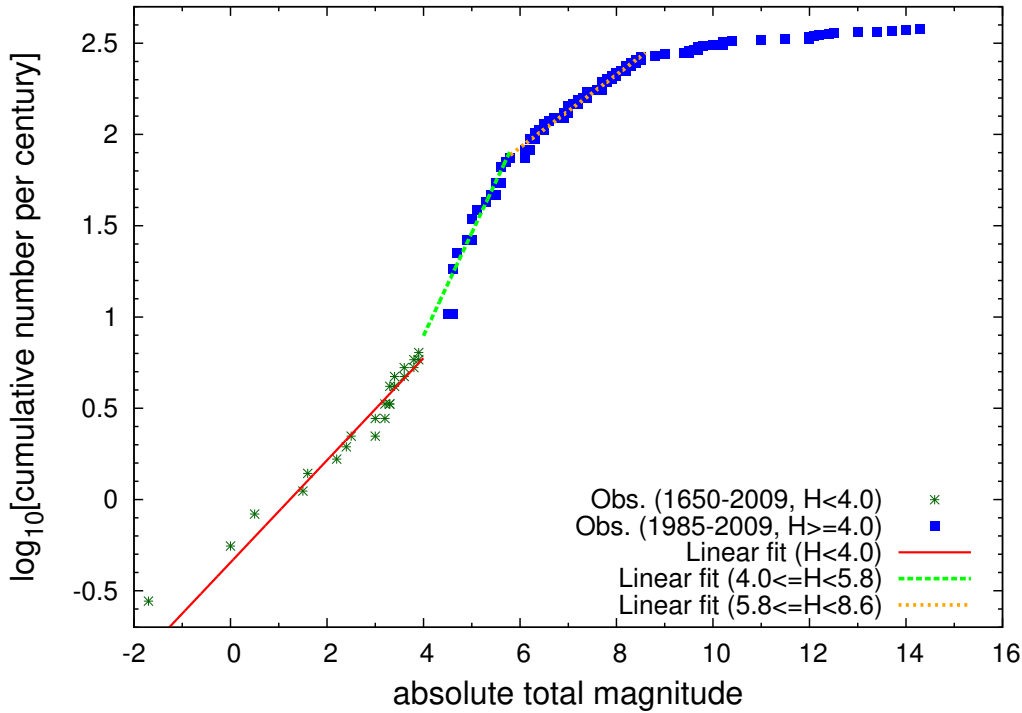}}
\caption{The cumulative distributions of $H$ for the sample of bright
  comets ($H < 4$) observed during the period 1650-2009, and the sample
  1985-2009 for comets with $H \geq 4$.}
\label{concat}
\end{figure}

We show in Fig. \ref{concat} the concatenated cumulative distributions
of $H$ for the sample of bright comets ($H < 4$) observed during 1650-2009,
and our assumed unbiased sample of LPCs with magnitudes $H \geq 4$ for
the period 1985-2009, both linear fits normalized to units of comets
century$^{-1}$. The concatenated distribution shows three segments: one
for the brightest comets ($H < 4$) with a slope 0.28, other for those
with intermediate brightness ($4 \leq H < 5.8$) with a slope 0.56,
and other for less bright comets ($5.8 \leq H < 8.6$) with a slope
0.20, as found in Sections 4.2 and 4.1, respectively. The cumulative
distribution tends to flatten for comets fainter
than $\sim 8.6$, and it levels off for $H \gsim 12$, in agreement with
what we said before about the scarcity of LPCs fainter than $H \simeq
12$.\\

We can convert the cumulative magnitude distribution law: $\log_{10}N_{<H} =
C + \alpha H$, into a cumulative size distribution (CSD) law, $N_{>R}$, by
means of the relation between the radius $R$ and $H$ given by equation
(\ref{rvh}). We obtain

\begin{equation}
N_{>R} \ = \ AR^{-s},
\label{s_dist}
\end{equation}
where $A$ is a normalization factor, the exponent $s = \alpha/b$, and
$b=0.13$ (cf. equation (\ref{rvh})). Likewise, from equation (\ref{rvh})
we can convert the magnitude ranges into ranges of $R$, as shown in Table
\ref{as_parameters}. We can also see in the table the values of the
parameters $A$ and $s$ obtained for the best-fit solutions. With the $A$
values of Table \ref{as_parameters} we obtain cumulative numbers
expressed in comets century$^{-1}$.\\

\begin{table}[h]
\centering
\caption{{\small Numerical values for the parameters $A$ and $s$ of
equation (\ref{s_dist})}}
\vspace{1mm}
\begin{scriptsize}
\begin{tabular}{|l r r|}
\hline
 Range of $R$ (km) & $A$ & $s$ \\ \hline
$R > 2.4$  & 38.8 & $2.15 \pm 0.75$ \\
$1.4 < R \leq 2.4$ & 329 & $4.31 \pm 0.80$ \\
$0.6 < R \leq 1.4$ & 130 & $1.54 \pm 0.15$ \\ \hline
\end{tabular}
\end{scriptsize}
\label{as_parameters}
\end{table}

As mentioned in the Introduction, there have been a few
attempts before to derive the magnitude distribution of LPCs, and in
some cases also their sizes. The knee in the $H$-distribution at $H
\simeq 6$ is a well established feature (e.g. Everhart 1967b, Sekanina
and Yeomans 1984). From the analysis of the magnitude distribution of
LPCs observed during 1830-1978, Donnison (1990) derived a slope
$\alpha \simeq 0.34$ for comets with $H < 6$. From the study of
Vsekhsvyatskii's sample, Hughes (2001) found a slope $\alpha \simeq
0.36$ within the range $3.5 \lsim H \lsim 6.6$. This corresponds to an
exponent $s = 2.77$ for the cumulative $R$-distribution, i.e. Hughes
obtained something in between our computed values of $\alpha$ and $s$
for our first and second leg ($H < 4$ and $4 \leq H < 5.8$). The
problem is that Hughes worked with LPCs of all perihelion
distances, thus bringing to his sample distant comets of very uncertain
computed values of $H$, and also more affected by observational
selection effects.\\

We can also compare our derived CSD for LPCs with those derived for other
populations. Tancredi et al. (2006) derived an exponent $s=2.7 \pm 0.3$
for a sample of Jupiter family comets (JFCs) with well determined nuclear
magnitudes and $q<2.5$ AU. On the other hand, Lamy et al. (2004)
derived a smaller value of $s \simeq 1.9 \pm 0.3$. A
recent re-evaluation of the CSD of JFCs by Snodgrass et el. (2011)
leads to a somewhat greater value of the slope: $s = 1.92 \pm 0.20$
applicable to comets with radii $\geq$ 1.25 km. The
sample of JFCs considered by these authors covered a range of radii from
sub-km to several km, i.e. it roughly overlaps part of our first leg of
brighter comets, the second and third leg, for which we derived
exponents of 2.15, 4.31 and 1.54 respectively.\\

More light can be shed from theoretical models. For instance, Dohnanyi
(1969) derived an exponent $s=2.5$ for a population in collisional
equilibrium. Kenyon and Bromley (2012) have considered a
protoplanetary disk divided in 64 annuli covering a range of distances
to the central star between 15-75 AU. Then the authors simulate the
evolution of a swarm of planetesimals distributed among the different
annuli in
order to follow a predetermined surface density law for the disk. The
population goes through a process of coagulation and fragmentation
leading to a few oligarchs (large embryo planets), and a large number
of small planetesimals that will largely evolve through destructive
collisions. What is suggestive for our study is that the authors find
an exponent $s=2$ for the CSD of the evolve population that remains in
the range $R \sim 10-100$ km, while the exponent is somewhat higher
than 2 in the range 1-10 km. Then, Kenyon and Bromley's (2012) results
match very well our computed value of $s=2.15$ for the brighter LPCs
with $H < 4$ ($R \gsim 2.4$ km). Furthermore, the population of larger
comets might precisely be the one that best preserves its primordial
size distribution. As we will see below, smaller LPCs may have gone
through recent phenomena upon approaching the Sun, as e.g. splitting
into two or more pieces, fading into meteoritic dust, that has greatly
changed its primordial distribution, as shown in Fig. \ref{concat}.

\section{New comets among the observed LPCs}

New comets are usually considered to be those with original energies
in the range $0 < E_{or} < 10^{-4}$
AU$^{-1}$\footnote{The usual convention is that the
  energies of elliptic orbits are negative though, for simplicity, in
  what follows we will take them as positive.}, which appear as a spike
in the $E_{or}$-histogram of LPCs, as first pointed out by Oort (1950).
Since the typical energy change by planetary perturbations is $>>
10^{-4}$ AU$^{-1}$, these comets are presumably new incomers in the
inner planetary region. Admittedly this may not be true in all
cases. By integrating the orbits of ``new'' comets backwards to their
previous perihelion passages, considering planetary perturbations and
the tidal force of the galactic disk, Dybczy\'nski (2001) found that
nearly 50\% of the so called new comets actually passed before by the
planetary region with $q < 15$ AU. From these results he proposed a
new definition of new comet based on the condition that the
perihelion distance of the previous passage had to be $q_{pre} > 15$
AU, thus discarding the
criterion based on the original energy. Yet Dybczy\'nski's definition
has its own shortcomings. Firstly, the computed $q_{pre}$ strongly
depends on the modeled galactic potential, and on the almost unknown
stellar perturbations. Furthermore, the boundary at $q=15$ AU to
discriminate between ``new'' and ``old'' comets is rather
arbitrary. By shifting this value upward or downward we can get
different new/old ratios. Therefore, we will stick to the classic
definition of new comet based on its original energy, on the dynamical
criterion that such a comet comes from the Oort cloud and, thus, has
been greatly influenced in its way in by the combined action of
galactic tidal forces and passing stars. Evolved comets will then be
defined as those with binding energies above $10^{-4}$ AU$^{-1}$, so
they are no longer influenced by external perturbers. They have
already passed before by the inner planetary region (interior to
Jupiter's orbit).

\subsection{The observed fraction of new comets in the incoming flux of
LPCs}

We considered in the first place our sample of LPCs for the period
1900-2009. Unfortunately,
reliable computed original energies are available for only a fraction
of them. Many comets, mainly those observed prior to about 1980, do
not have computed values of $E_{or}$. Therefore, we also analysed the more
restricted -though more complete- sample of LPCs brighter than $H = 9$
discovered in the last quarter of century (1985-2009). Following other
authors (Francis 2005, Neslu\v{s}an 2007), we also considered more
restricted samples from sky surveys that are presumably less biased,
and that have computed $E_{or}$ for most of their members. Since our
main interest here is to derive the ratio new-to-evolved LPCs, and not
so much their absolute magnitudes, for the sky surveys we considered
comets covering a much wider range of perihelion distances ($0 < q <
4$ AU) under the assumption that up to $q \simeq 4$ AU the detection
probability was quite high. In short, we analysed the following
samples:

\begin{itemize}
\item LPCs with $P > 10^3$ yr and $q < 1.3$ AU observed during the
  period 1900-2009 (232 comets).
\item LPCs with $P > 10^3$ yr, $q < 1.3$ AU, and brighter than $H = 9$
  for the period 1985-2009 (68 comets).
\item LPCs with $P > 10^3$ AU and $q < 4$ AU discovered by LINEAR (73
  comets).
\item LPCs with $P > 10^3$ AU and $q < 4$ AU discovered by other large
  sky surveys (Siding Spring, NEAT, LONEOS, Spacewatch, Catalina) (45
  comets).
\end{itemize}

The computed original energies were taken from Marsden and Williams's
(2008) catalogue, complemented with some values from Kinoshita and
Nakano electronic catalogues
(see references in the Introduction). For our comet sample 1900-2009, the
values of $E_{or}$ are also included in Tables \ref{sample_modern}
and \ref{sample_recent}.\\

\begin{figure}[h]
\centerline{\includegraphics[width=8cm]{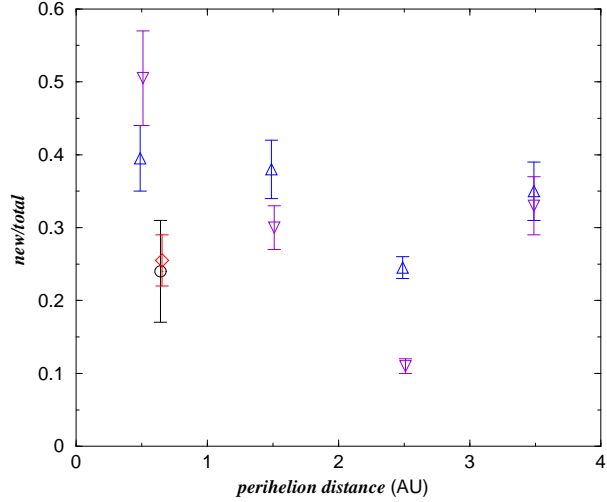}}
\caption{The ratio of new/total LPCs for the different samples as a
function of the average perihelion distance of the considered range.
The samples are: LPCs observed during the period 1900-2009 with
$q < 1.3$ AU and all
magnitudes (diamond); LPCs observed during 1985-2009 with $q < 1.3$
AU and $H < 9$ (circle); LINEAR comets with $q < 4$ AU discriminated in
intervals of 1 AU (triangle down); LPCs from other sky surveys
discriminated in intervals of 1 AU (triangle up).}
\label{new_total}
\end{figure}

Let $n_{LPC} = n_{new} + n_{ev}$ be the total number of LPCs,
comprising both new and evolved LPCs which are given by the quantites
$n_{new}$ and $n_{ev}$, respectively. From the $n_{new}/n_{LPC}$ ratios
found for the different samples, as shown in Fig. \ref{new_total}, we
found as an average

\begin{equation}
\frac{n_{new}}{n_{LPC}} \simeq 0.3 \pm 0.1,
\label{new_lpc}
\end{equation}
or $n_{new}/n_{ev} \sim 0.43$, namely, for about 7 evolved LPCs with
$P > 10^3$ yr we have approximately 3 new comets.\\

The error bars of the derived values shown in Fig. \ref{new_total}
take into account only the uncertainty within the considered sample
due to LPCs of unknown or poorly determined $E_{or}$. However, it does
not consider the uncertainty inherent to the finite size $n$ of our
samples of random elements (that goes as $n^{1/2}$). How
the different error sources play is a complex matter, but we still consider
that the uncertainty associated to, or lack of computed values of $E_{or}$,
may be the principal one.

\subsection{The theoretically expected new/evolved LPCs ratio as a
function of the maximum number of passages}

Let us assume that we have an initial population of $n_{new}$ comets
injected in orbits with $q < 1.3$ AU. About half of this population
will be lost to the interstellar space, and about half will return
as evolved LPCs, in general with bound energies $E > 10^{-4}$
AU$^{-1}$. After the second passage, these returning comets can either be
ejected or come back with a different bound energy, and the same
process will repeat again in the following passages. The energies that
bound comets get in successive passages can be described as a random
walk in the energy space, in which in every passage a given comet
receives a kick in its energy due to planetary perturbations. After
$N$ passages the number $n_N$ that will remain bound to the solar
system is (e.g. Fern\'andez 1981, 2005)

\begin{equation}
n_N \sim \frac{1}{2} n_{new} N^{-1/2}.
\end{equation}

The total number of returns of the $n_{new}$ comets as evolved LPCs,
$n_{ev}$, can be computed by summing the returning comets between $N=1$
and a maximum number of passages $N_{max}$ that is set either by the
physical lifetime of the comet, or by the dynamical timescale
(expressed in number of revolutions) to reach an orbit with $P < 10^3$
yr (or a binding energy $E > 10^{-2}$ AU$^{-1}$), namely

\begin{equation}
n_{ev} = \sum^{N_{max}}_{N=1} n_N \sim \frac{1}{2} n_{new}
\sum^{N_{max}}_{N=1} N^{-1/2} \sim n_{new} N_{max}^{1/2}.
\label{nev}
\end{equation}

Since $n_{ev}/n_{new} \simeq 7/3$, from equation (\ref{nev}) we find that

\begin{equation}
N_{max} \simeq \left(\frac{n_{ev}}{n_{new}}\right)^2 \sim 5.5.
\end{equation}
This result is in good agreement with that found by Wiegert and
Tremaine (1999) from numerical simulations of fictitious Oort cloud
comets injected into the planetary region with different fading laws.
These authors obtained as the best match to the observed distribution
of orbital elements a survival of roughly 6 orbits for the 95\% of
the comets, while the remainder 5\% did not fade. We may argue that
this 5\% of more robust comets are associated with the largest members
of the comet population.\\

The random walk in the energy space, where each step has a typical
change $\Delta E_t \sim 10^{-3}$ AU$^{-1}$, allows us to estimate the
typical number of passages $N_{dyn}$ required to reach an energy
$E > E_L = 10^{-2}$ AU$^{-1}$ (or a period $P < 10^3$ yr)

\begin{equation}
N_{dyn} \sim \left(\frac{E_L}{\Delta E_t}\right)^2 = 10^2.
\end{equation}

We find that $N_{max} << N_{dyn}$, i.e. most comets will be destroyed by
physical processes (sublimation, outbursts, splittings) before
reaching energies $ > E_L$.

\subsection{Comets coming from the outer and from the inner Oort cloud}

Among the new comets, we distinguish in turn those
coming form the outer Oort cloud, with energies in the range $0 <
E_{or} < 0.3$, from those coming from the inner Oort cloud, with
energies in the range $0.3 \leq E_{or} < 1$ (both in units of $10^{-4}$
AU$^{-1}$). Outer Oort cloud comets (semimajor axis $a \gsim
3.3 \times 10^4$ AU) can be driven from the outer planetary region to
the inner planetary region in a single revolution by the combined
action of galactic tidal forces and passing stars, so they can overshoot
the powerful Jupiter-Saturn gravitational barrier (e.g. Fern\'andez 2005,
Rickman et al. 2008). On the contrary, the stronger gravitationally
bound comets in the inner Oort cloud will require more than one
revolution to diffuse their perihelia to the inner planetary region,
so they will meet in their way in the Jupiter-Saturn barrier, with the
result that most of the comets will be ejected before reaching the
near-Earth region. From the study of the discovery conditions
of a sample of 58 new comets discovered during 1999-2007,
Fern\'andez (2009) found
that comets in the energy range $0 < E_{or}(10^{-4}\mbox{ AU$^{-1}$}) <
0.3$ show a uniform $q$-distribution, as expected from comets
injected straight into the inner planetary region from a thermalised
population, while the $q$-distribution of comets with original
energies $0.3 < E_{or}(10^{-4}\mbox{ AU$^{-1}$}) < 1$ show an increase
with $q$ which may be attributed to the Jupiter-Saturn barrier that
prevents most of the inner Oort cloud comets from reaching the inner
planetary region. As a corollary, we infer that the ratio outer-to-inner
Oort cloud comets derived for the Earth neighbourhood should decrease
when we consider new comets beyond Jupiter.\\

We note than when we talk about comets {\it coming} from a given
region, we mean the region attained during the last orbit. As shown by
Kaib and Quinn (2009), a comet from the inner Oort cloud whose perihelion
approaches the Jupiter-Saturn barrier can receive a kick in its energy that
sends it to the outer Oort cloud where the stronger galactic tidal
forces and stellar perturbations can deflect it to the near-Earth
region, overcoming in this way the Jupiter-Saturn barrier. Therefore,
we cannot tell for sure if a comet has resided for a long time in
the outer or inner Oort cloud, but only the place from where it comes
in the observed apparition. It is very likely that Kaib and Quinn's
mechanism provides a steady leaking of comets from the inner to the
outer Oort cloud.\\

The estimate of the ratio outer-to-inner is a very complex
matter since we are dealing with very narrow energy ranges ($\Delta E
\simeq 0.3 \times 10^{-4}$ AU$^{-1}$ for the outer Oort cloud, and
$\Delta E \simeq 0.7 \times 10^{-4}$ AU$^{-1}$ for the inner Oort cloud),
so the errors in the computation of original orbital energies may be of
the order of these ranges. Even though the formal errors of the
computed original orbital energies of comets of quality classes 1A and
1B in Marsden and Williams's (2008) catalogue are $\pm 5$ and $\pm 12$
(in units of $10^{-6}$ AU$^{-1}$),
respectively, unaccounted nongravitational (NG) effects may shift the
computed $E_{or}$ by several tens of units, as shown by Kr\'olikowska
and Dybczy\'nski (2010). We will neglect NG forces for the moment, and
come back to this complex issue below. In
Table \ref{in_out} we show the outer-to-inner ratio of QC 1A comets.
We consider three ranges of perihelion distances: $0 < q \leq 1.5$ AU,
$1.5 < q \leq 3.0$ AU, $3.0 < q \leq 4.5$ AU. The available samples
are very likely incomplete, though we hope that the incompleteness
factor is similar for comets from the inner and from the outer Oort
cloud, so the ratio will remain more or less constant.\\

\begin{table}[h]
\centering
\caption{{\small Ratio of comets coming from the outer Oort cloud to
    those coming from the inner Oort cloud.}}
\vspace{1mm}
\begin{scriptsize}
\begin{tabular}{|c|c|}
\hline
 $\Delta q$ & Outer : Inner \\ \hline
$0 < q \leq 1.5$ &  7 : 5 \\
$1.5 < q \leq 3$ & 10 : 8 \\
$3 < q \leq 4.5$ & 13 : 18 \\ \hline
\end{tabular}
\end{scriptsize}
\label{in_out}
\end{table}

There is a slight predominance of new comets coming from the outer
Oort cloud than from the inner Oort cloud in the first two $q$
ranges. In the most distant one the situation reverses and there are
more comets coming from the inner Oort cloud. As said above, the
outer-to-inner ratio does not have to keep constant throughout the
planetary region. On the contrary, as $q$ increases it is expected
that more comets from the inner Oort cloud will be present as we
pass over and leave behind the Jupiter-Saturn barrier.
We have also checked the outer-to-inner ratios for the sample of QC 1B
comets despite their lower quality. We find 9:9, 6:6 and 1:4 for
the ranges $0 < q \leq 1.5$ AU, $1.5 < q \leq 3$ AU, and $3 < q
\leq 4.5$ AU, respectively. The trend remains more or less the same:
we may argue that the slight decrease in the outer-to-inner ratios from
1A to 1B comets for the first two ranges, $0 < q \leq 1.5$ AU and
$1.5 < q \leq 3$ AU, is due to some blurring in the computed
original orbit energies of the lower quality class 1B comets. If there
is some reshuffling, it will be more likely that errors will put
an outer Oort Cloud comet in the energy range of inner Oort Cloud
comets than the other way around, simply because the width of the inner Oort
Cloud energy range is more than twice that of the outer Oort Cloud.\\ 	

Let us call $n_{in}$ and $n_{out}$ the number of new comets
coming from the inner and outer Oort cloud, respectively, such that
$n_{new} = n_{in} + n_{out}$. From the previous analysis, we find
that $n_{out}$ may be a little above $n_{in}$ in the zone closer to the
Sun (say $q \lsim 3-4$ AU) with an error bar that may leave the lower
end slightly below one, so we can estimate

\begin{equation}
\frac{n_{out}}{n_{in}} \simeq 1.1 \pm 0.2
\label{outer_inner}
\end{equation}

Let us now come back to the problem of unaccounted NG forces that
might affect the $E_{or}$ values computed by Marsden and Williams
(2008). Kr\'olikowska and Dybczy\'nski (2010) and Dybczy\'nski and
Kr\'olikowska (2011) have recomputed the orbits of ``new'' comets, as
defined by Marsden and Williams, including NG terms in the equations
of motion. The authors find that the inclusion of NG forces tends to
shift the computed $E_{or}$ to greater values (smaller semimajor
axes). This shift could affect the computed ratio $n_{out}/n_{in}$, as
some comets will move from original hyperbolic orbits to the Oort
cloud, and others from the Oort cloud to evolved orbits. Kr\'olikowska
and Dybczy\'nski (2010) computed original orbits of comets with
$E_{or} < 10^{-4}$ AU$^{-1}$ and $q < 3$ AU, whereas Dybczy\'nski and
Kr\'olikowska (2011) considered those with $q > 3$ AU. While the
orbits computed with NG forces were the best for the first case, for
comets with $q > 3$ AU only 15 out of 64 comets presented measurable
NG forces. Altogether, they assembled a sample of 62 comets whose
computed original energies (mostly NG solutions) fall in the range $0
< E_{or} < 10^{-4}$ AU$^{-1}$. From these computed set of energies, we
find that 26 comets come from the outer Oort cloud ($E_{or} < 0.3 \times
10^{-4}$ AU$^{-1}$), and 36 from the inner Oort cloud. If we limit the
sample to comets with $q > 3$ AU, less affected by NG forces, the
numbers are: 20 from the outer Oort cloud and 18 from the inner Oort
cloud, i.e. an $n_{out}/n_{in}$ ratio slightly above unity. If we now
extrapolate this result to smaller $q$, we should expect a slight
increase of this ratio (see discussion above), though still compatible
with the one shown in equation (\ref{outer_inner}).

\section{Evolutive changes in the size distribution: sublimation and
  splitting}

Comet splitting is a rather common phenomenon observed in comets
(e.g. Chen and Jewitt 1994). In most cases small chunks and debris are
released of very short lifetime, so the process can be described as a
strong erosion of the comet nucleus surface, but essentially it remains
as a single body. Yet, in some
occasions the splitting of the parent comet may lead to two or more
massive fragments that become unbound, and may last for several
revolutions, thus producing daughter comets
that may be discovered as independent comets. We have
several cases of comet pairs among the observed LPCs which are shown in
Table \ref{split_comets}. We include in the table only those comets
whose splittings are attributed to endogenous causes, namely we are
leaving aside comets that tidally split in close encounters with the
Sun or planets.\\

\begin{table}[h]
\centering
\caption{{\small LPC pairs.}}
\vspace{1mm}
\begin{scriptsize}
\begin{tabular}{|l c|}
\hline
Pair & $q$ (AU) \\ \hline
1988 F1 Levy and 1988 J1 Shoemaker-Holt & 1.17 \\
1988 A1 Liller and 1996 Q1 Tabur & 0.84 \\
2002 A1 LINEAR and 2002 A2 LINEAR & 4.71 \\
2002 Q2 LINEAR and 2002 Q3 LINEAR & 1.31 \\ \hline
\end{tabular}
\end{scriptsize}
\label{split_comets}
\end{table}

The splitting phenomenon is a consequence of the very fragile nature
of the comet material. For instance, from the tidal breakup of comet
D/1993 F2 (Shoemaker-Levy 9) in a string of fragments, Asphaug and
Benz (1996) found that the tidal event could be well modeled by
assuming that the nucleus was a strengthless aggregate of grains.
The high altitudes at which fireballs (of probable
cometary origin) are observed to disrupt also lead to low strengths
between $\sim 10^3$ - $10^5$ or $10^6$ erg cm$^{-3}$ (Ceplecha and
McCrosky 1976; Wetherill and ReVelle 1982). Also from the analysis
of the height of meteoroid fragmentation, Trigo-Rodr\ii guez and
Llorca (2006) find that cometary meteoroids have typical strengths
of $\sim 10^4$ erg cm$^{-3}$, though it is found to be of only
$\sim 4 \times 10^2$ erg cm$^{-3}$ for the extremely fluffy particles
released from 21P/Giacobini-Zinner.\\

Samarasinha (2001) has proposed an interesting model to explain the
breakup of small comets like C/1999 S4 (LINEAR). The author assumes a
rubble-pile model for the comet nucleus with a network of
interconnected voids. The input energy comes from the Sun. The solar
energy penetrates by conduction raising the temperature of the outer
layers of the nucleus. As the heat wave penetrates, an
exothermic phase transition of amorphous ice into cubic ice may occur
when a temperature of 136.8 K is attained, releasing an energy
of $8.4 \times 10^8$ erg g$^{-1}$ (Prialnik and Bar-Nun 1987).
The released energy can go into the sublimation of some water ice
and the liberation of the CO molecules, trapped in the ice matrix,
that propagate within the network filling the voids, thus building a
nucleus-wide gas pressure able to disrupt a weakly consolidated
body. As the self-gravity scales as $R^2$, larger comet nuclei might be
able to hold the fragments together. Therefore, we may infer that
there is a critical radius below which the breakup with dispersion of
fragments occurs, since the self-gravity is too weak to reassemble the
fragments.\\

\begin{figure}[h]
\centerline{\includegraphics[width=14cm]{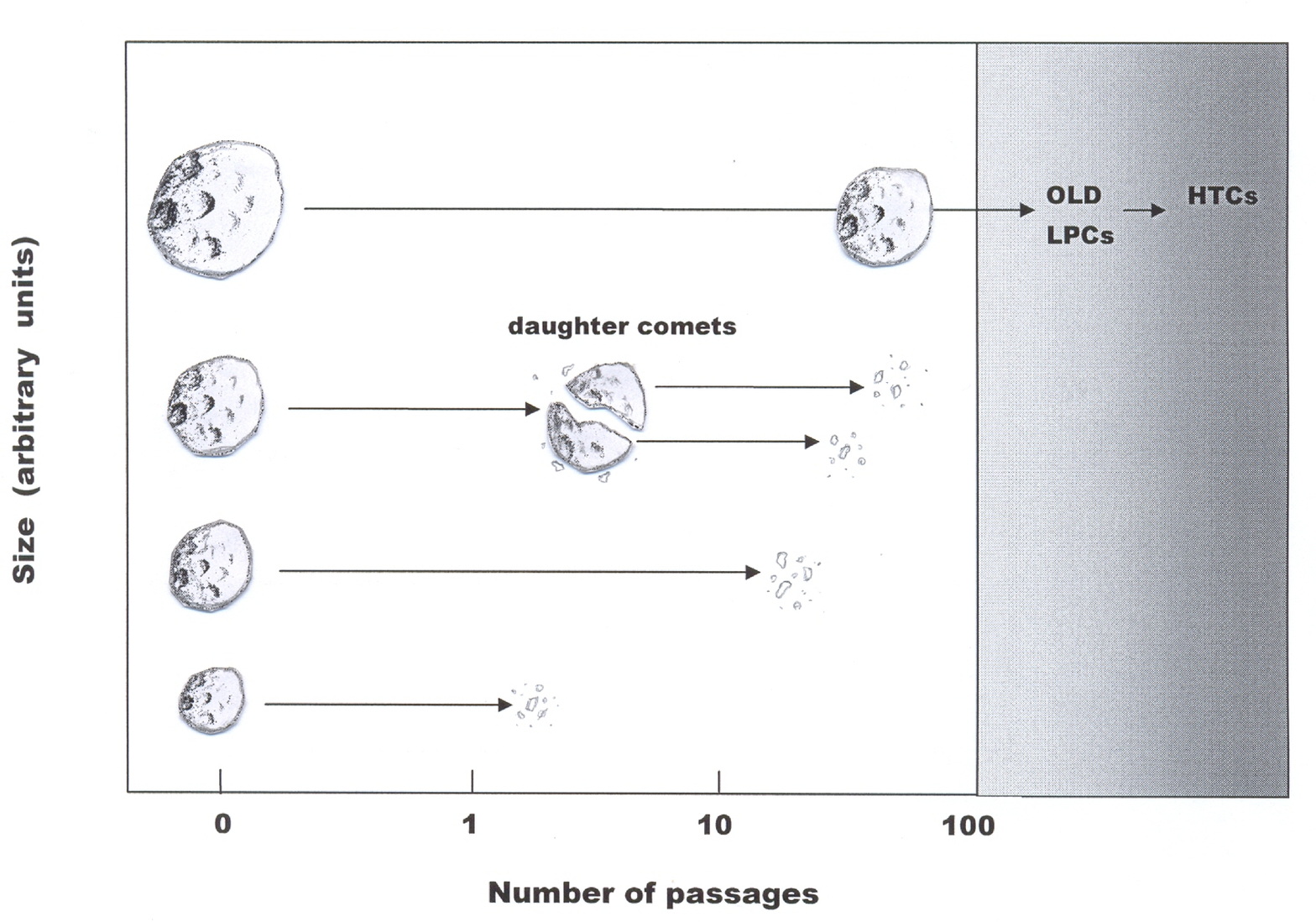}}
\caption{Sketch depicting the physical evolution of nuclei of
  different sizes.}
\label{evol}
\end{figure}

Fig. \ref{evol} depicts possible physical pathways for comets of
different sizes that summarize what we have discussed in this paper. Large
comets (about ten-km size or larger) can survive for hundreds or thousands
of revolutions so they may reach old dynamical ages (Halley types). Large
comets are essentially lost by hyperbolic ejection. Medium-size comets
(several km) does not have enough gravity field to avoid the separation
of fragments upon breakups which may lead to the production of daughter
comets that continue their independent lives until disintegration or
ejection. Small comets (about one km) can last several
passages until disintegration without producing daughter comets (namely
fragments are too small to last for long enough to be detected as
independent comets). Finally, very small comets (some tenths km)
quickly disintegrate after one or a few passages at most.

\section{A numerical model}

We developed a simple model to simulate the dynamical and physical
evolution of cometary nuclei of different sizes entering in the inner
Solar System from the Oort Cloud. Our aim was to try to reproduce, in a
qualitative sense and broad terms, the observed size distribution (as
inferred from the absolute magnitude distribution of the
observed LPCs shown in Fig. \ref{concat}). We performed numerical
simulations for large samples of fictitious comets with initial
parabolic orbits (original orbital energies $E=0$), random inclinations,
and a perihelion distance $q = 1$ AU, varying the nuclear
radius from 0.5 km up to 50 km, with a size bin of 0.25 km, so we
considered 198 initial radii. For every initial radius, we computed
samples of $10^4$ fictitious comets, so we studied a total of $198
\times 10^4$ comets for each one of the runs (that have associated the
sets of initial
conditions shown in Table \ref{init_cond}). In every passage of a
test comet by the planetary region, we compute the orbital energy
change $\Delta E$ due to planetary perturbations. The perihelion
distance and the angular orbital elements were assumed to remain
constant through the simulation, as they are little
affected by planetary perturbations. We assumed that $\Delta E$
followed a random Gaussian
distribution (e.g. Fern\'andez 1981), with a mean value = 0 and a
standard deviation $\sigma_E = 7 \times 10^{-4}$ AU$^{-1}$. We note
that the only purpose for assuming a ``random inclination'' for the
comet's orbit is to adopt an inclination-averaged value of $\sigma_E$.
The simulations were terminated when the test comets reached one of
the following end states:

\begin{itemize}
\item they became periodic, i.e. they reached an orbital energy $\ge
10^{-2}$ AU$^{-1}$, corresponding to a orbital period $P \le 10^3$ yr;
\item they were ejected from the Solar System, i.e. they reached
(in our convention of sign) a negative orbital energy;
\item they were disintegrated after several passages due to sublimation,
i.e reached a radius below a certain \emph{minimum radius} $R_{min} = 0.25$
km; or
\item they reached a maximum number of 2000 orbital revolutions.
\end{itemize}

The following model parameters were allowed to change in each simulation:

\begin{itemize}
\item the radius decrease $\Delta R$ per perihelion passage, due to
sublimation and other related effects, as for instance outbursts and
release of chunks of material from the surface. We adopted for the
radius decrease the following expression: $\Delta R = \lambda\Delta R_s$,
where $\lambda$ is a dimensionless factor, and $\Delta R_s$ is the
decrease in the radius due to sublimation of water ice, which for a
nucleus of bulk density $\rho$ is given by

\begin{equation}
\Delta R_s = \frac{\Delta m_s}{\rho},
\label{drs}
\end{equation}
where $\Delta m_s$ is the mass loss per unit area due to sublimation.
This quantity can be computed from the polynomial fit by Di Sisto et al.
(2009) to theoretical thermodynamical models of a free-sublimating
comet nucleus

\begin{equation}
\Delta m_s \approx 1074.99 -4170.89q+8296.96q^2-8791.78q^3+4988.9q^4
-1431.4q^5+162.975q^6,
\label{dms}
\end{equation}
where $q$ is given in AU and $\Delta m_s$ in g cm$^{-2}$. We solve
equation (\ref{dms}) for $q=1$ AU, that was our adopted perihelion
distance for the test comets, and took $\rho = 0.4$ g cm$^{-3}$ for
computing $\Delta R_s$ from equation (\ref{drs}).

\item a lower and an upper limit radii ($R_{sp1}$ and $R_{sp2}$,
respectively) for the occurrence of splitting leading to the creation
of two daughter comets. In other words, if the comet reached
a radius within the range [$R_{sp1}$ $R_{sp2}$], it was allowed to split
in a pair of comets of a half the mass of the parent comet each, with a
certain frequency $f_{sp}$.
\item the frequency of splittings $f_{sp}$.
\end{itemize}

In some simulations, we added a few more parameters: an intermediate
critical radius $R_{spi}$, between $R_{sp1}$ and $R_{sp2}$, and two
frequencies ``high'' and ``low'', $f_{sph}$ and $f_{spl}$, respectively,
instead of $f_{sp}$. Under these conditions, comets with radii
$R_{sp1} < R < R_{spi}$ were allowed to split with a frequency
$f_{sph}$, and comets with $R_{spi} < R < R_{sp2}$ with a frequency
$f_{spl}$.\\

We show in Table \ref{init_cond} the initial conditions chosen for our
four runs of $198 \times 10^4$ fictitious comets each.\\

\begin{table}[h]
\centering
\caption{{\small Initial conditions}}
\vspace{1mm}
\begin{scriptsize}
\begin{tabular}{|c|r r r r r r r r|}
\hline
 Run & $\Delta R$ & $\lambda$ & $R_{sp1}$ & $R_{spi}$ & $R_{sp2}$ & $f_{sp}$ &
$f_{sph}$ & $f_{spl}$  \\
       & (km)& & (km) & (km) & (km)&  &  &   \\ \hline
1 & 0.025& 7.7 & 1.6& - & 2.7& 0.1 & - & - \\
2 & 0.025& 7.7 & 1.8& - & 5.0& 0.05& - & - \\
3 & 0.0125& 3.9 & 1.8& 3.4&5.0 & - & 0.1& 0.05 \\
4 & 0.00625& 1.9 & 1.8& 3.4& 5.0& - & 0.1& 0.05 \\
 \hline
\end{tabular}
\end{scriptsize}
\label{init_cond}
\end{table}

\begin{figure}[h]
\centerline{\includegraphics[width=14cm]{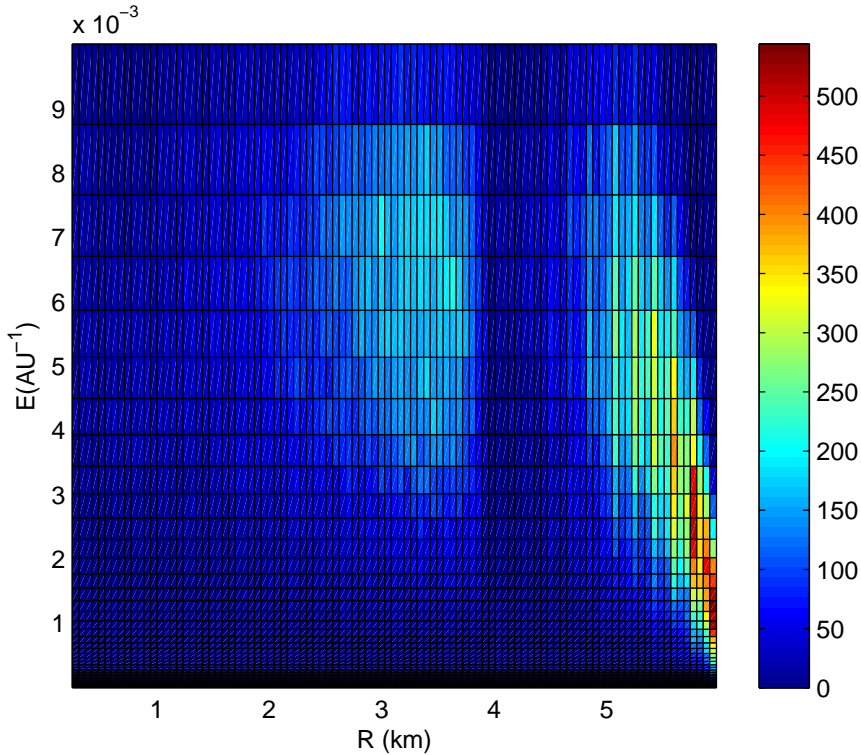}}
\caption{Physico-dynamical evolution of a sample of $10^4$ fictitious
comets starting in parabolic orbits of radius $R = 6$ km taken
from the Run 2. The two diffusion zones correspond to the parent and
daughter comets. The different colours indicate different number
densities of passages with given sets ($R$,$E$).}
\label{map}
\end{figure}

Fig. \ref{map} illustrates the physico-dynamical evolution of one
of our samples of $10^4$ comets with an initial radius $R = 6$ km
taken from Run 2. All the comets are injected in parabolic orbits
($E=0$) (lower right corner of the panel). The survivors return in
orbits of different bound energies $E$ and decreasing radii $R$ due
to erosion. The evolution can be seen as a diffusion in the parametric
plane $(R,E)$. The different colours represent the different number
of passages with a given combination of $R$ and $E$. As the comets
get dynamically older (a greater average number of passages), their
radii decrease by erosion. This is the reason why the diffusion
proceeds from the lower right corner to the upper left side of the
diagram.\\

When the model comets decreased their radii below $R_{sp2} = 5$ km,
they were allowed to split in two daughter comets with a frequency of
one every 20 passages. The splitting event is random, so this was
simulated by picking a random number $z$ within the interval $(0,1)$,
and imposing the condition that the splitting occurred if $z <
0.05$. The production of daughter comets gives rise to a second wave
of comet passages toward the left-hand side of the diagram. Comets
that reached radii below $R_{sp1} = 1.8$ km were assumed to proceed to
disintegration without producing daughter comets.\\

Once we computed the 198 samples of different initial radii $R_i$
of a given run, we assembled the $R_i$ samples into a single comet
population. We next tried to match the differential $R$-distribution
of this population to an assumed
differential radius distribution $n_R = dN_R/dR \propto R^{-\nu}$,
where we adopted $\nu = s + 1 = 3.15$, namely the same index as
that derived for the largest comets (cf. Section 4.3), that we
assumed for our model as the representative of the Oort cloud
population. In other words, we assume that the largest observed LPCs have
been preserved almost unscathed since their injection into the inner
planetary region, so their observed CSD reflects that of Oort cloud
comets, and that it extends to smaller comets in the Oort cloud, down
to the smallest radius considered in our model.
Since all the samples have $10^4$ comets (and thus
provide an uniform differential $R$-distribution), we transformed it
to an $R$-distribution $\propto R^{-\nu}$ just by multiplying a given
sample of radius $R_i$ by the scaling factor $R_i^{-\nu}$. By adding
the different samples of $R_i$, scaled by $R_i^{-\nu}$, we can
obtain the cumulative distribution for the sample of new + evolved
comets.\\

\begin{figure}[h]
\centerline{\includegraphics[width=14cm]{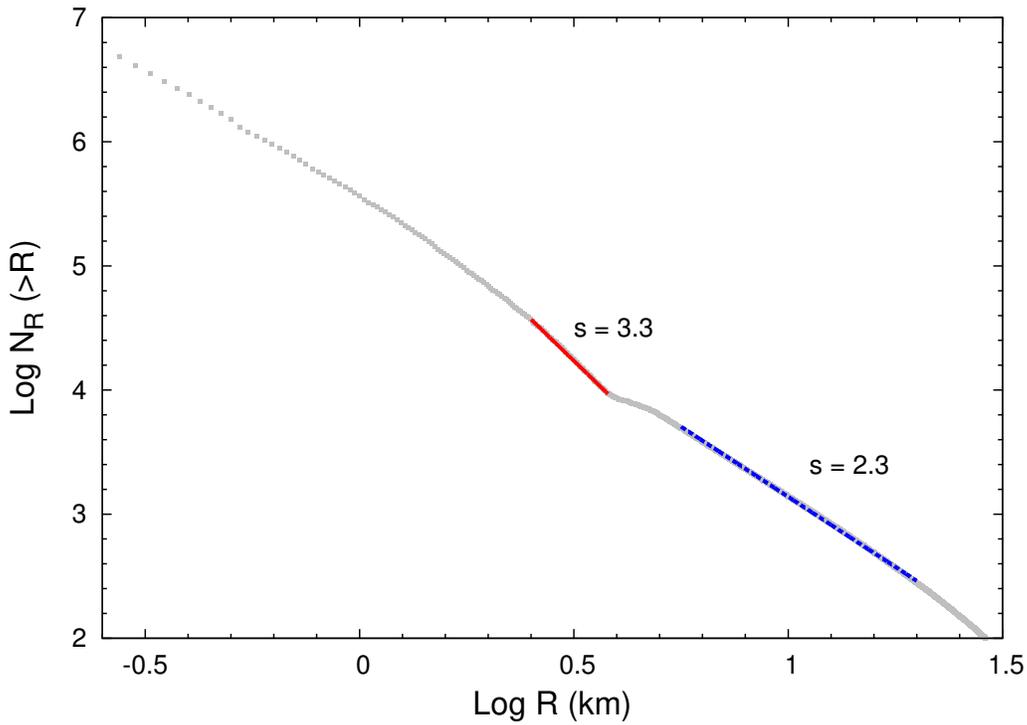}}
\caption{Computed cumulative size-distribution of a sample of $198
\times 10^4$ fictitious comets of Run 2.}
\label{simul2}
\end{figure}

Fig. \ref{simul2} shows the cumulative distribution of the nuclear
radius (evolved as well as new comets), with both axes in logarithmic
scales, corresponding to the Run 2. The model parameter values for
this simulation were: $\Delta R = 0.025$ km ($\lambda \simeq 7.7$),
$f_{sp}$ = 1/20, $R_{sp1} = 1.8$ km, and $R_{sp2} = 5.0$ km. A size
bin of 0.025 km was used.
We can see that the slope of the largest comets
$= +2.3$ is not very different from the primordial one ($+2.15$).
Yet, in the region where daughter comets are created the slope
raises to $+3.3$. The model slope is still lower than the one derived
before ($+4.31$) for the range of radii between 1.4 - 2.4 km, which
suggests that splitting may not be the only cause for the change of
slope at $H \sim 4$, and that other cause (primordial?) may add to the
previous one.
We obtained a ratio $n_{new}$/$n_{LPC}= 0.22$ for this simulation.
The computed ratio is still lower than the observed one (0.30, cf.
equation (\ref{new_lpc})), which suggests that we still require
higher erosion rates per orbital revolution: $\Delta R > 0.025$ km,
or $\lambda > 7.7$, to match the observed ratio $n_{new}/n_{LPC}$.
This ratio decreased to about 0.07 for comets with radii $R >
2.75$ km, which is to be expected since large comets have very
likely longer physical lifetimes, thus yielding more passages as
evolved comets per new comet.\\

As expected, the match between the computed and the observed ratios
$n_{new}$/$n_{LPC}$ worsened when we considered lower erosion rates.
We obtained $n_{new}$/$n_{LPC}$ = 0.16 when we changed to
$\Delta R = 0.0125$ km ($\lambda \simeq 3.9$), while the slopes remain
almost unchanged.
For $\Delta R = 0.00625$ km ($\lambda \simeq 1.9$), $R_{sp1}$ = 1.8 km,
$R_{spi}$ = 3.4 km, $f_{sph}$ = 0.1, and $f_{sl}$  = 0.05, we obtained
$n_{new}$/$n_{LPC}$ = 0.12, again with slopes close to -3.3 and -2.3.
As seen, the results for the cumulative $R$-distribution are quite
robust, almost independent of the adopted erosion rate. On the other
hand, the differences in the computed ratio $n_{new}$/$n_{LPC}$ are
quite substantial from run to run.

\section{Summary and conclusions}

The most important results of our work can be summarised in the
following points:

\begin{enumerate}
\item We have ellaborated an updated catalogue of absolute total
  visual magnitudes $H$ for the LPCs with $q < 1.3$ AU observed during
  1970-2009.

\item We have analysed the cumulative distribution of $H$, finding at
  least a three-modal distribution with slopes $\alpha \simeq 0.28 \pm
  0.10$ for the brightest comets with $H < 4$, $\alpha \simeq 0.56 \pm
  0.10$ for comets with intermediate brightness ($4 \leq H < 5.8$), and
  $\alpha \simeq 0.20 \pm 0.02$ for the fainter comets ($5.8 \leq H <
  8.6$). From the relation between $H$ and the radius $R$
  (cf. equation (\ref{rvh})), we can derive the cumulative size
  distribution, which can be expressed by a power-law: $N_{>R}
  =AR^{-s}$, where $s \simeq +2.15$ for $R \gsim 2.4$ km, $s \simeq +4.31$
  for $1.4 \lsim R \lsim 2.4$ km, and $s \simeq +1.54$ for $0.6 \lsim R
  \lsim 1.4$ km.

\item The change at $H \sim 4$ from a rather shallow slope to a steep
  one may be at least partially explained as a result of splitting and
  separation of the fragments, thus leading to the creation of two or
  more daughter comets. Comets brighter than $H \sim 4$ may as well
  split but their gravitational fields are strong enough to keep the
  fragments bound.

\item Comets fainter than $H \sim 5.8$ may be too small to survive for
  more than one or a few passages, so parent comets as well as their
  daughters might go through a fast fading process, thus explaining
  the flattening of the cumulative $H$-distribution.

\item The cumulative $H$-distribution flattens even more for LPCs
  fainter than $H \simeq 8.6$, reaching a ceiling at $H \sim 12$
(diameter $\sim 0.5$ km). We
  suggest that the scarcity of extremely faint LPCs is a real
  phenomenon, and not due to observational selection effects. This is
  supported by several sky surveys which have been very successful at
  discovering a large number of very faint NEAs (absolute magnitudes
  $\sim 14-25$) in cometary orbits, but failed to discover a
  significant number of faint LPCs.

\item The fraction of new comets within the LPC population with
$q < 1.3$ AU is found to be $0.3 \pm 0.1$, namely we have about 3
new comets for every 7 evolved ones. This implies that the average
  number of returns of a new comet coming within 1.3 AU from the Sun
  is about 2.3.

\item The ratio between new comets coming from the outer Oort cloud to
  those coming from the inner Oort cloud, within 1.3 AU from the Sun,
  is found to be: $1.1 \pm 0.02$.

\item We have simulated the physical and dynamical evolution of LPCs
  by means of a simple numerical model. We find that erosion rates
  greater than about 8 times the free sublimation rate of water ice
are required to match the observed new-to-evolved LPCs ratio. With a
  splitting rate of one every 20 revolutions, we find an increase in
  the slope for intermediate brightness comets from 2.3 to 3.3. Even
  though there is an agreement in qualitative terms, the computed
  increase in the slope falls short of reproducing the observed slope:
  4.31 for comets of this size range. Therefore, other effects might
  be at work to explain such an increase as, for instance, primordial
  causes (namely related to the accretion processes and collisional
  evolution), or the production of multiple fragments that might
  survive several revolutions as independent comets. As suggested by
Stern and Weissman (2001), the scattering of cometesimals by the
Jovian planets to the Oort cloud was preceded by an intense collision
process between cometesimals and their debris. This process could have
led to a heavily fragmented population of comets, just in the size
range of a few km. Comets with radii $R \gsim 5-10$ km could have
suffered catastrophic collisions, but their gravitational fields
were powerful enough to reaccumulate the majority of their fragments, thus
preserving their primordial size distribution, though with an internal
structure like a 'rubble pile' (e.g. Weissman 1986). If the initial
comet population with radii in the range $\sim 1-10$ km was
collisionally relaxed, then the slope of the CSD was close to $s =
2.5$ (e.g. Dohnanyi 1969, Farinella and Davis 1996).
Yet, since bodies with $R \gsim$ a few km reaccumulated their fragments,
the primordial CSD of slope $s \simeq 2.15$ might have been
preserved, thus reflecting the coagulation/fragmentation conditions in
the early protoplanetary disk (Kenyon and Bromley 2012). Therefore we
might argue that when the scattered
cometesimals reached the Oort cloud, they already had a bimodal size
distribution. Breakups into daughter comets during passages into
the inner planetary region might only enhance an already existing
bimodality in the size distribution. No doubt, this is a point that
deserves further study.
\end{enumerate}

\bigskip

{\large \bf ACKNOWLEDGMENTS}

\bigskip

We want to thank Daniel Green for providing us data in electronic form
from the ICQ archive. We also thank Ramon Brasser and the referee,
Hans Rickman, for their comments and criticisms on an earlier version
of the manuscript that greatly helped to improve the presentation of
the results. AS acknowledges the national research agency,
Agencia Nacional de Investigaci\'on e Innovaci\'on, and the
Comisi\'on Sectorial de Investigaci\'on Cient\ii fica, Universidad
de la Rep\'ublica, for their financial support for this work, which is
part of her PhD thesis at the Universidad de la Rep\'ublica, Uruguay.

\end{document}